\documentclass[aps,twocolumn,pra,superscriptaddress,amsmath,showpacs,tightenlines]{revtex4-1}
\usepackage{amssymb}
\usepackage{amsmath}
\usepackage{graphicx}
\usepackage{subfigure}
\usepackage{natbib}
\usepackage{epsfig}
\usepackage{amsfonts}
\usepackage{mathrsfs}
\usepackage{xcolor}
\usepackage[toc,page,title,titletoc,header]{appendix}

\begin{document}

\title{Phase controlled single-photon nonreciprocal transmission in a one-dimensional waveguide}
\author{Zhihai Wang}
\affiliation{Center for Quantum Sciences and School of Physics, Northeast Normal University, Changchun 130024, China}
\author{Lei Du}
\affiliation{Center for Quantum Sciences and School of Physics, Northeast Normal University, Changchun 130024, China}
\author{Yong Li}
\email{liyong@csrc.ac.cn}
\affiliation{Beijing Computational Science Research Center, Beijing 100193, China}
\affiliation{Center for Quantum Sciences and School of Physics, Northeast Normal University, Changchun 130024, China}
\author{Yu-xi Liu}
\affiliation{The Institute of Microelectronics, Tsinghua University, Beijing 100084, China}
\affiliation{Center for Quantum Information Science and Technology, BNRist, Beijing 100084, China}

\begin{abstract}
We study the controllable single-photon scattering via a one-dimensional waveguide which is coupled to a two-level emitter and a single-mode cavity simultaneously. The emitter and the cavity are also coupled to each other and  form a three-level system with cyclic transitions {within the zero- and single-excitation subspaces}. As a result, the phase of emitter-cavity coupling strength serves as a sensitive control parameter.  When the emitter and cavity locate at the same point of the waveguide, we demonstrate the Rabi splitting and quasidark-state--induced perfect transmission for the incident photons. More interestingly, when they locate at different points of the waveguide, a controllable nonreciprocal transmission can be realized {and the non-reciprocity is robust to the weak coupling between the system and environment.} Furthermore, we demonstrate that our
theoretical model is experimentally feasible with currently available technologies.
\end{abstract}
%\date{\today}

\maketitle
\section{Introduction}
Optical nonreciprocal devices allow the propagation of photons from one side to be superior than that from the opposite side. Due to its potential applications in quantum sensing and  information process, the nonreciprocal signal transmission has been studied widely in various of physical systems, such as the opto-mechanical or {electro-mechanical systems}~\cite{opto1,opto2,opto3,opto4,opto5,opto6,opto7,
opto8,opto9,opto10,opto11,opto12},  parity-time-symmetry optical systems~\cite{PT1,PT2,PT3,PT4,PT5,PT6}, cavity QED systems~\cite{nonl1,nonl2,nonl3,nonl4,nonl5,nonl6,nonl7,nonl8,nonl9,nonl10,nonl11,nonl12,nonl13} and atomic systems~\cite{atomic1,atomic2,atomic3,atomic4,atomic5,atomic6}. On the other hand, the controllable photon transmission in quantum network composed by the waveguide and quantum node plays a central role in the design of quantum transistors~\cite{trans1,trans2,trans3,trans4,trans5,trans6}, quantum routers~\cite{trans7,trans8}, and frequency converters~\cite{trans9,trans10,trans11,trans12}. Therefore, it is a natural issue to study the photon nonreciprocal transmission and controllability of non-reciprocality in quantum network.

One of the physical principles behind the nonreciprocal transmission is the breaking of time-reversal symmetry. But it is actually not common in the traditional waveguide QED system, in which the waveguide is coupled to other systems (e.g., atoms, quantum dots). Because the linear size of the atom is considered to be much smaller than the wavelength of the light, and the Hamiltonian usually possesses time-reversal symmetry~\cite{trans1,trans2,trans3,trans4,trans5,trans6,trans7,trans8,trans9,
trans10,trans11,trans12}. However, in the recent experimental studies, the coupling between superconducting artificial atoms~\cite{gaint1,gaint2}(e.g. transmon~\cite{gaint3}) and the surface acoustic waves~\cite{gaint4,gaint5} has been realized successfully. In this system, the transmon serves as a ``giant atom'', which is connected to the waveguide at multiple points~\cite{gaint6,gaint7,giant8}. In this situation, the time-reversal symmetry is naturally broken, making it possible to realize the nonreciprocal transmission.

Motivated by the above achievements, we here consider an emitter-cavity interacting system, forming a Jaynes-Cummings model~\cite{JC}, which is coupled to a waveguide with linear dispersion relation. We investigate the single-photon scattering in such a system, where the emitter and the cavity locate at one or two different points of the waveguide, depending on the available physical realization. In our system, the phase of the emitter-cavity coupling strength, which cannot be eliminated by any gauge transformation, plays as a sensitive controller for the scattering behavior. When the emitter and the cavity locate at the same point of the waveguide, we show the Rabi splitting and the quasidark-state--induced~\cite{ZW,Tian} perfect transmission in the scattering spectrum. When they locate at different points of the waveguide, we demonstrate the phase-controlled nonreciprocal transmission, and the underlying physical mechanism is further revealed in the momentum space. {Our preliminary studies here also show that the time-reverse symmetry breaking induced nonreciprocal transmission is robust to the emitter-cavity detuning and their interaction to the surrounding environments.}

The rest of the paper is organized as follows. In Sec.~\ref{model}, we show the Hamiltonian and formulate the single-photon scattering process. In Sec.~\ref{one}, we discuss the phase control to the scattering process with the emitter and the cavity locating at the same point of the waveguide. In Sec.~\ref{two}, we discuss the controllable nonreciprocal transmission when they locate at two different points of the waveguide. {In Sec.~\ref{ds}, we briefly discuss the effects of the emitter-cavity detuning, the emitter-cavity distance, and the interaction to the environments on the single-photon transmission.} In Sec.~\ref{proposal}, we propose an experimental scheme based on the superconducting quantum circuits to realize the controllable single photon nonreciprocal transmission.  Finally, we give a conclusion in Sec.~\ref{con}. {In the Appendix, we give the expression of the Hamiltonian in the momentum space.}

\section{Model and single-photon scattering}
\label{model}

\begin{figure}
\begin{centering}
\includegraphics[width=1\columnwidth]{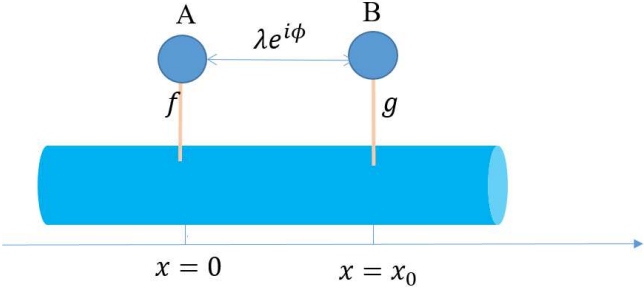}
\par\end{centering}
\caption{{Schematic diagram for two quantum nodes $A$ and $B$ which are coupled to a waveguide at the points $x=0$ and $x=x_0$, respectively. The two quantum nodes are also coupled to each other directly.}}
\label{scheme}
\end{figure}

{We consider a system which is schematically shown in Fig.~\ref{scheme}.  A one-dimensional waveguide, which serves as a quantum channel, which is coupled to two quantum nodes $A$ and $B$, located at $x=0$ and $x=x_0$, respectively. The two quantum nodes are also coupled to each other directly. The quantum nodes can be composed by spins, cavities, or atoms, etc. However, in our following studies, we will take the quantum node $A$ as a single-mode cavity and node $B$ as a two-level emitter as an example.}

In our consideration, the Hamiltonian $H$ of the system is composed of three parts, i.e., $H=H_{s}+H_{w}+V$. The first part is
\begin{equation}
H_{s}=\omega_{a}a^{\dagger}a+\Omega|e\rangle\langle e|+\lambda(e^{i\phi}a^{\dagger}\sigma_{-}+h.c.),\label{eq:hs}
\end{equation}
which is the Hamiltonian of the emitter coupled to the cavity.
Hereafter, we set $\hbar=1$. $a$ is
the annihilation operator of the single-mode cavity field with frequency
$\omega_{a}$. $\Omega$ is the transition frequency from the ground
state $|g\rangle$ to the excited state $|e\rangle$ of the two-level
emitter, and $\sigma_{+}=(\sigma_{-})^{\dagger}=|e\rangle\langle g|$
is the raising operator. The real numbers $\lambda$ and $\phi$ are
the magnitude and the phase of the coupling constant between
the two-level emitter and the cavity field. For the sake of simplicity, we will assume
that the cavity field resonantly interacts with the two-level emitter, i.e., $\omega_a=\Omega=\omega$. {The effect of the emitter-cavity detuning will be discussed in Sec.~\ref{ds}.}

The second part $H_w$ of the Hamiltonian $H$ is the free Hamiltonian of the waveguide. When the frequency of the emitter or/and the cavity is far away from the cut-off frequency of the dispersion relation, the Hamiltonian $H_w$ of the waveguide can be written as~\cite{trans3}
\begin{equation}
H_{w}=\int dx\{-iv_{g}c_{R}^{\dagger}(x)\frac{d}{dx}c_{R}(x)+iv_{g}c_{L}^{\dagger}(x)\frac{d}{dx}c_{L}(x)\},
\label{inH}
\end{equation}
where $v_{g}$ is the group velocity of the travelling photons in the waveguide. Here, we have set the length of the waveguide to be unity, so the group velocity $v_g$ has the same unit with the frequency. $c_{R}^{\dagger}(x)$ $(c_{L}^{\dagger}(x))$ is the bosonic creation operator for a right-going (left-going) photon at the
position $x$.

The third part $V$ of the Hamiltonian describes the interaction between the
waveguide and the two-level emitter as well as the cavity field. Under the rotating wave approximation, the Hamiltonian $V$ can be written as

\begin{eqnarray}
V & = & f\int dx\delta(x)[a^{\dagger}c_{R}(x)+a^{\dagger}c_{L}(x)+h.c.]\nonumber \\
 & + & g\int dx\delta(x-x_{0})[\sigma_{+}c_{R}(x)+\sigma_{+}c_{L}(x)+h.c.],
 \label{VV}
\end{eqnarray}
where $f$ ($g$) is the coupling strength between the cavity (emitter)
and the waveguide, and has been assumed as real numbers.  The Dirac-delta
function in the Hamiltonian $V$ implies that the cavity locates at
the point $x=0$, and the emitter locates at the point $x=x_0$. Below, we will discuss the cases of $x_0=0$ and $x_0\neq0$, respectively. {We emphasize that, besides the parameter $\phi$, the properties of the system are also sensitive to the phase $kx_0$, where $k$ is the wave vector of photon in the waveguide. To clarify this fact more clearly, we furthermore give the expression of Hamiltonian in the momentum space in the Appendix. }

We now consider that a single photon with wave vector $k$
is incident from the left side of the waveguide. Since the excitation
number in the system conserves, the eigenstate in
the single-excitation subspace can be written as
\begin{eqnarray}
|E_{k}\rangle & = & \int dx[\phi_{R}(x)c_{R}^{\dagger}(x)+\phi_{L}(x)c_{L}^{\dagger}(x)]|G\rangle\nonumber \\
 &  & +u_{e}\sigma_{+}|G\rangle+u_{a}a^{\dagger}|G\rangle,
\end{eqnarray}
where $|G\rangle$ represents that both the waveguide and the cavity are in the vacuum while the emitter is in the ground state $|g\rangle$. $\phi_{R}(x)$ and $\phi_{L}(x)$ are, respectively, the single-photon wave functions of the right-going and left-going modes in the waveguide. $u_{e}$ is the excitation amplitude of the emitter and $u_{a}$ is the excitation amplitude of the cavity field. Solving the stationary Sch$\rm{\ddot{o}}$dinger equation $H|E_{k}\rangle=E|E_{k}\rangle,$ we obtain the equations for the amplitudes as

\begin{subequations}
\begin{eqnarray}
-iv_{g}\frac{d}{dx}\phi_{R}(x)+f\delta(x)u_{a}+g\delta(x-x_{0})u_{e} & = & E\phi_{R}(x),\nonumber \\
\\
iv_{g}\frac{d}{dx}\phi_{L}(x)+f\delta(x)u_{a}+g\delta(x-x_{0})u_{e} & = & E\phi_{L}(x),\nonumber \\
\\
\omega u_{a}+\lambda u_{e}e^{i\phi}+f[\phi_{R}(0)+\phi_{L}(0)] & = & Eu_{a},\\
\omega u_{e}+\lambda u_{a}e^{-i\phi}+g[\phi_{R}(x_{0})+\phi_{L}(x_{0})] & = & Eu_{e}.
\end{eqnarray}
\label{amp}
\end{subequations}
We are now aiming to study the scattering behavior when a single photon with wave vector $k$ is incident from the left side of the waveguide. Therefore, the wave function $\phi_{R}(x)$ and $\phi_L(x)$ can be written in the forms of
\begin{eqnarray}
\phi_{R}(x) & = & e^{ikx}\{\theta(-x)+A[\theta(x)-\theta(x-x_{0})]+t\theta(x-x_{0})\},\nonumber \\
\label{R}\\
\phi_{L}(x) & = & e^{-ikx}\{r\theta(-x)+B[\theta(x)-\theta(x-x_{0})]\},\label{L}
\end{eqnarray}
with
\begin{equation}
\theta(x)=\begin{cases}
1 & x>0\\
\frac{1}{2} & x=0\\
0 & x<0
\end{cases}.
\end{equation}
The underlying physics behind the above ansatz for the left-going
and right-going photons can be explained as follows. The right-going
photons incident from the regime of $x<0$ can be transmitted or
reflected when it arrives at the cavity-waveguide connecting point
at $x=0$ with the reflection and transmission amplitudes, denoted
by $r$ and $A$ respectively. The transmitted photon will freely
propagate until it meets the emitter-waveguide connecting point at
$x=x_{0}$, and may be reflected or transmitted
with the amplitudes $B$ and $t$, respectively. Substituting the photon
amplitudes in Eqs.~(\ref{R}) and (\ref{L}) into Eqs.~(\ref{amp}a) to (\ref{amp}d), we obtain the transmission amplitude $t$ as ($E=v_gk$)

\begin{equation}
t=\frac{\Delta^{2}-\lambda^{2}-2fg\lambda e^{-i\phi}\sin (kx_{0})/v_{g}}{\Delta^{2}-\lambda^{2}+iK},\label{eq:tt}
\end{equation}
where $\Delta=E-\omega$ is the detuning between the incident photon
and the cavity field (emitter), and
\begin{equation}
K=\frac{\Delta(f^{2}+g^{2})+2e^{ikx_{0}}fg[\lambda\cos\phi+fg\sin (kx_{0})/v_{g}]}{v_{g}}.
\end{equation}

\section{Phase control with one connecting point}
\label{one}
In the above section, we have obtained the amplitude $t$ of the single-photon
transmission. For the sake of simplicity, we here discuss the situation
of $x_{0}=0$, that is, the two-level emitter and the cavity locate at the same point
of the waveguide. We will discuss the case of $x_{0}\neq0$
in the next section.

\begin{figure}
\begin{centering}
\includegraphics[width=1\columnwidth]{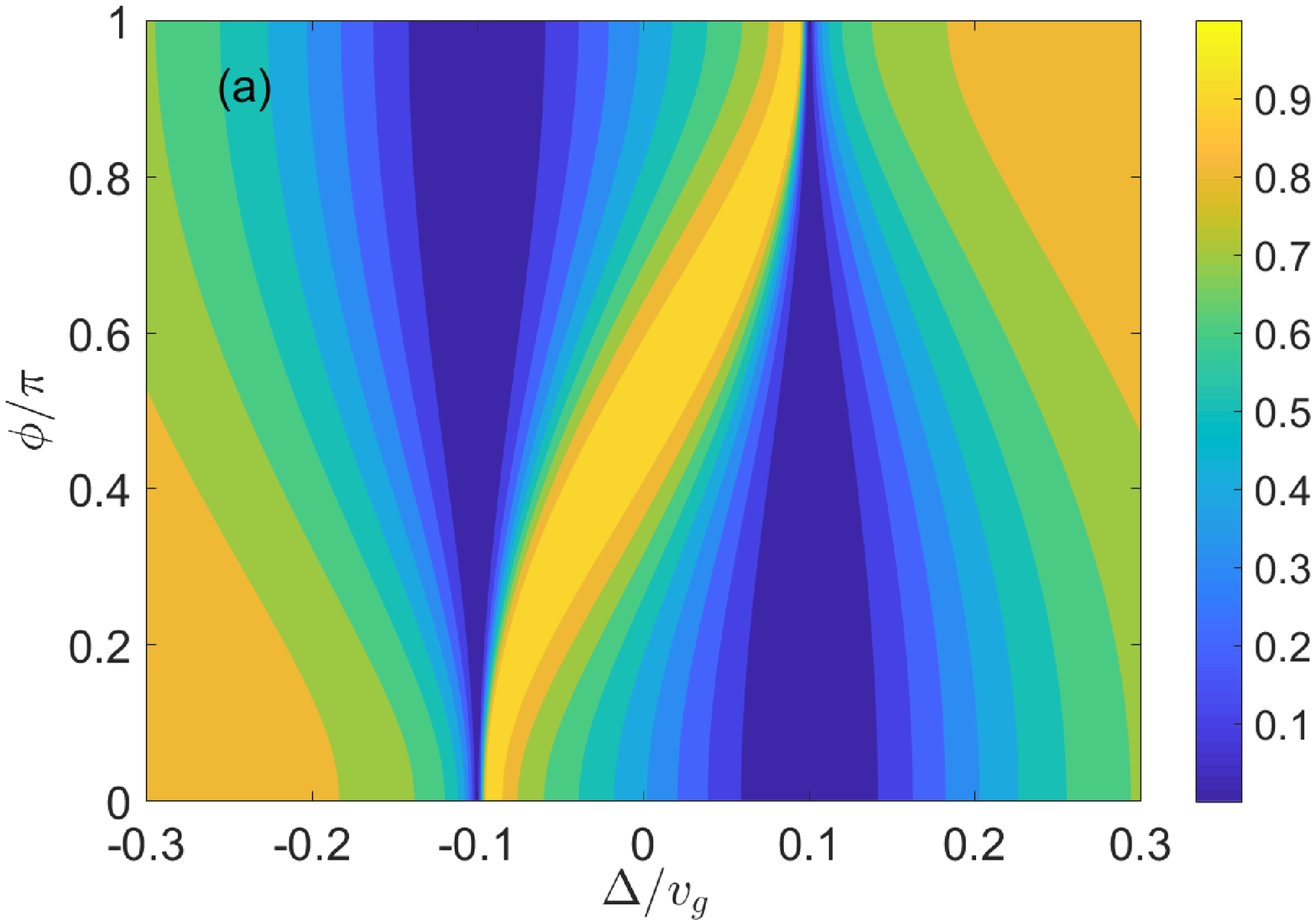}
\par\end{centering}
\begin{centering}
\includegraphics[width=1\columnwidth]{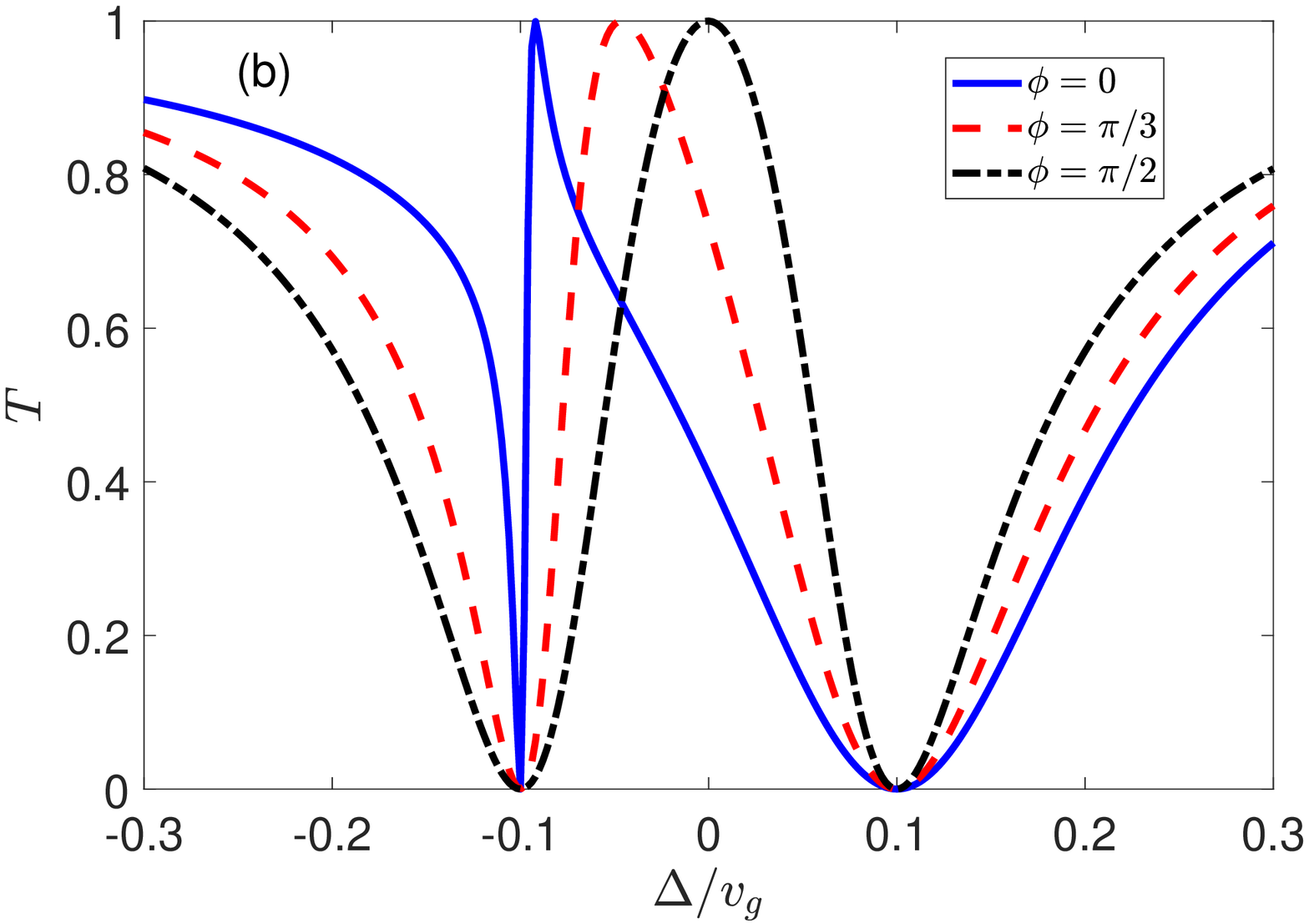}
\par\end{centering}
\caption{ (a) The transmission rate $T$ as functions of $\phi$ and $\Delta$. (b) The transmission rate $T$ as a function of $\Delta$ for given $\phi$.
The parameters are set as $(\lambda,f,g)=(0.1,0.3,0.2)v_g$. Here, we consider the case that the cavity and the emitter are connected to the waveguide at the same point, that is, $x_0=0$. }
\label{trans}
\end{figure}

For $x_{0}=0$, the transmission amplitude $t$ in Eq.~(\ref{eq:tt})
can be simplified to
\begin{equation}
t=\frac{\Delta^{2}-\lambda^{2}}{\Delta^{2}-\lambda^{2}+i[\Delta(f^{2}+g^{2})+2fg\lambda\cos\phi]/v_{g}}.\label{eq:tx0}
\end{equation}
In Fig.~\ref{trans}(a), we plot the transmission rate $T=|t|^{2}$ as a function
of the detuning $\Delta$ and the phase $\phi$. The dependence
of $T$ on the phase $\phi$ can be explained from the energy-level
diagram of the total system. As shown in Fig.~3, the photon in the waveguide will induce the transition between the states $|0;g\rangle$ and $|1;g\rangle$ or between the states $|0;g\rangle$ and $|0;e\rangle$ of the emitter-cavity system with the coupling strength $f$ or $g$, while the direct coupling between the emitter and the cavity with the coupling strength $\lambda e^{i\phi}$ induces the transitions between the states $|1;g\rangle$ and $|0;e\rangle$. These three transitions form a cyclic type configuration. Therefore, the total phase $\phi$ cannot be eliminated by any canonical
gauge, and it affects the scattering behavior naturally as shown in
Fig.~\ref{trans}. Actually, the similar closed cyclic energy-level diagram can
also be found in many other systems, such as superconducting artificial
atoms~\cite{cy1,cy2}, chiral molecules~\cite{cy3,cy4,cy5,cy6}, cavity-QED systems~\cite{cy7}, and cavity optomechanical
systems~\cite{cy8,cy9}, in which the phase control to quantum phenomenon has attracted
much attention.

\begin{figure}
\begin{centering}
\includegraphics[width=1\columnwidth]{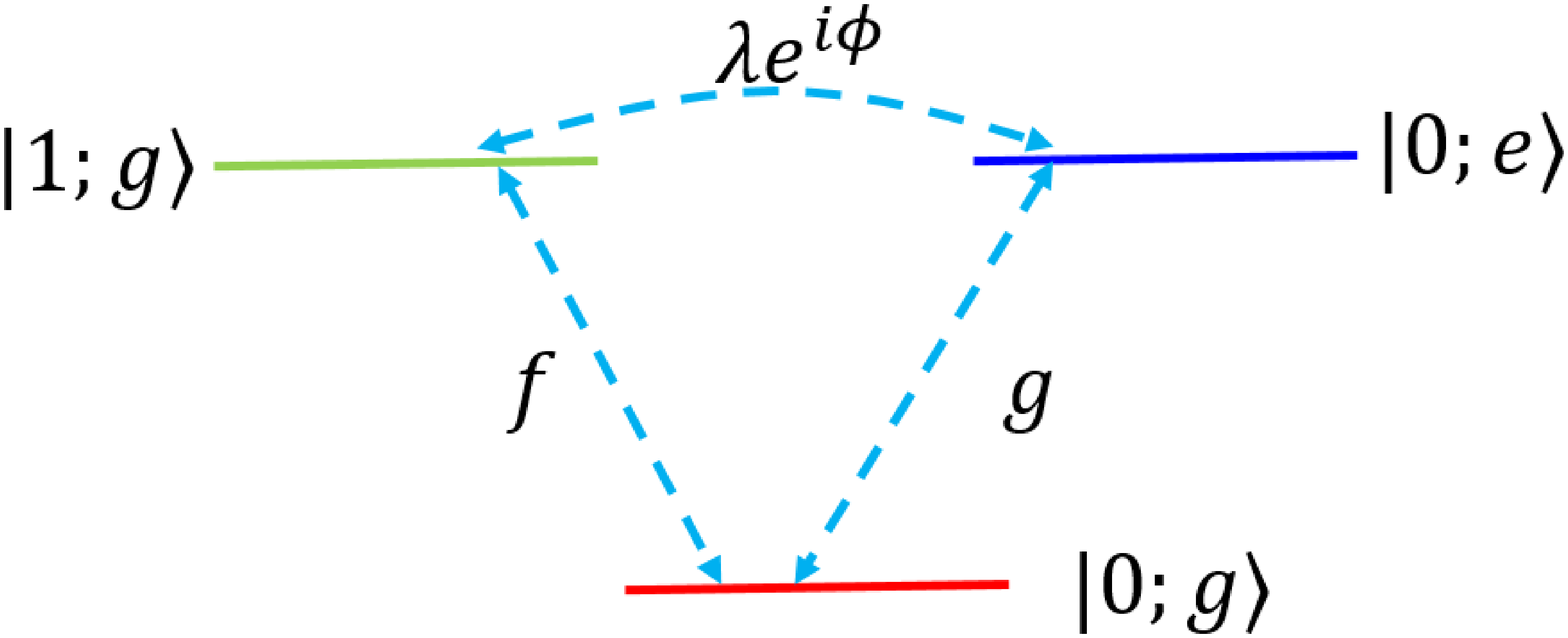}
\par\end{centering}
\caption{The cyclic energy-level diagram formed by the coupling between the two-level emitter and single-mode cavity and their interaction with the waveguide. Here, we give the energy-level diagram up to single-excitation subspace.}
\label{circle}
\end{figure}
The transmission rate $T$ as a function of the detuning $\Delta$
for different $\phi$ is plotted in Fig.~\ref{trans}(b). It shows that the complete
reflection ($T=0$) occurs whenever the incident photon is resonant with the two
dressed states of $H_{s}$ in the single-excitation subspace, that is, $E=\omega+\lambda$ or $E=\omega-\lambda$. These two dressed states are
\begin{eqnarray}
|\psi_{+}\rangle & = & \frac{1}{\sqrt{2}}(e^{i\phi}|0;e\rangle+|1;g\rangle),\\
|\psi_{-}\rangle & = & \frac{1}{\sqrt{2}}(-e^{i\phi}|0;e\rangle+|1;g\rangle),
\end{eqnarray}
{which satisfy $H_{s}|\psi_{\pm}\rangle=(\omega\pm\lambda)|\psi_{\pm}\rangle$.
Actually, the incident photons pass through the waveguide in two ways. One is to be directly transmitted without interacting with the emitter-cavity system. The other one is absorbed by the emitter-cavity system and induce the transitions between the states of $|0;g\rangle$ and two dressed states. The absorbed photon will be re-emitted via the interaction between the emitter-cavity system and the waveguide, and propagates to the left and right sides of the waveguide with the same probability. When the incident photon is just resonant with the $|0;g\rangle\leftrightarrow |\psi_+\rangle$ or $|0;g\rangle\leftrightarrow |\psi_-\rangle$ transition frequencies,  the direct transmitted and right-going re-emitted photon  will
cancel each other via the destructive interference process, leading to a perfect reflection. As a result, the incident photon with frequencies $E=\omega\pm\lambda$
($\Delta=\pm\lambda$) will not be transmitted and this is the vacuum Rabi splitting phenomenon.}

Furthermore, we also observe that a transmission peak ($T=1$) appears between
the two valleys, the corresponding frequency is phase-dependent and
can be obtained by virtue of  Eq.~(\ref{eq:tx0}) as
\begin{equation}
\Delta(f^{2}+g^{2})+2fg\lambda\cos\phi=0,
\end{equation}
which yields
\begin{equation}
E=\omega-\frac{2fg\lambda\cos\phi}{f^{2}+g^{2}}.
\end{equation}

Physically speaking, the transmission peak associates with the
``quasidark-state'' which is similar to that in Refs.~\cite{ZW,Tian}. In the current
system, the waveguide supplies a common bath~\cite{ZW,clerk} for the emitter and cavity,
and the normalized quasidark state is expressed as
\begin{equation}
|D\rangle=\frac{f|e;0\rangle-g|g;1\rangle}{\sqrt{f^{2}+g^{2}}}.
\end{equation}
It can be checked that
\begin{equation}
\langle D|H_{s}|D\rangle=\omega-\frac{2fg\lambda\cos\phi}{f^{2}+g^{2}},
\end{equation}
and
\begin{equation}
V|D\rangle\otimes|\emptyset\rangle=0,
\label{v0}
\end{equation}
where $|\emptyset\rangle$ represents all of the modes in the waveguide are in their vacuum states. $V$ is given by Eq.~(\ref{VV}) and we have considered the case with $x_0=0$.  It is implied in Eq.~(\ref{v0}) that the incident photon with frequency $E=\omega-(2fg\lambda\cos\phi )/(f^{2}+g^{2})$ will not interact with the emitter-cavity system at all, and therefore will be completely transmitted.

In the above discussions, we have restricted to the situation that the emitter and the cavity locate at the same point of the waveguide. Here the time-reversal symmetry is kept, and it is reciprocal for the single-photon scattering. That is, the transmission rate of the left-going photons is equal to that of the right-going photons.

\section{Controllable nonreciprocal transmission with two different connecting points}
\label{two}

In this section, let us consider the situation of $x_{0}\neq0$, that is,
the emitter and the cavity locate at the different points of the waveguide.
Similar to the case of $x_{0}=0$, the phase $\phi$ also
plays a role as a sensitive controller to the single-photon scattering due to the cyclic energy-level diagram as discussed in Sec.~\ref{one}.
Meanwhile, it will show a nonreciprocal scattering behavior
for $x_{0}\ne0$, that is, when the single photon
is incident from the right side, the transmission rate will be different
from that when it is incident from the left side.

Compared to the case that the incident photons are from the left side, the opposite-propagation (that is, the incident photons are from the right side) is equivalent to exchange the locations of the cavity and emitter in the waveguide.
That is, the cavity locates at the point $x=x_0$ of the waveguide while the emitter
locates at the point $x=0$ of the waveguide.  On the other hand, we note that the single-photon scattering behavior only depends on the relative position between the points where the emitter and the cavity locate, and the absolute positions are actually meaningless. Therefore, the corresponding transmission amplitude when the incident photons are from the right side can be directly given by changing $x_0$ in Eq.~(\ref{eq:tt}) into $-x_0$, which yields
\begin{equation}
t'=\frac{\Delta^{2}-\lambda^{2}+2fg\lambda e^{-i\phi}\sin (kx_{0})/v_{g}}{\Delta^{2}-\lambda^{2}+iK'}
\label{tp}
\end{equation}
with
\begin{equation}
K'=\frac{\Delta(f^{2}+g^{2})+2e^{-ikx_{0}}fg[\lambda\cos\phi-fg\sin (kx_{0})/v_{g}]}{v_{g}}.
\end{equation}

\begin{figure}
\begin{centering}
\includegraphics[width=1\columnwidth]{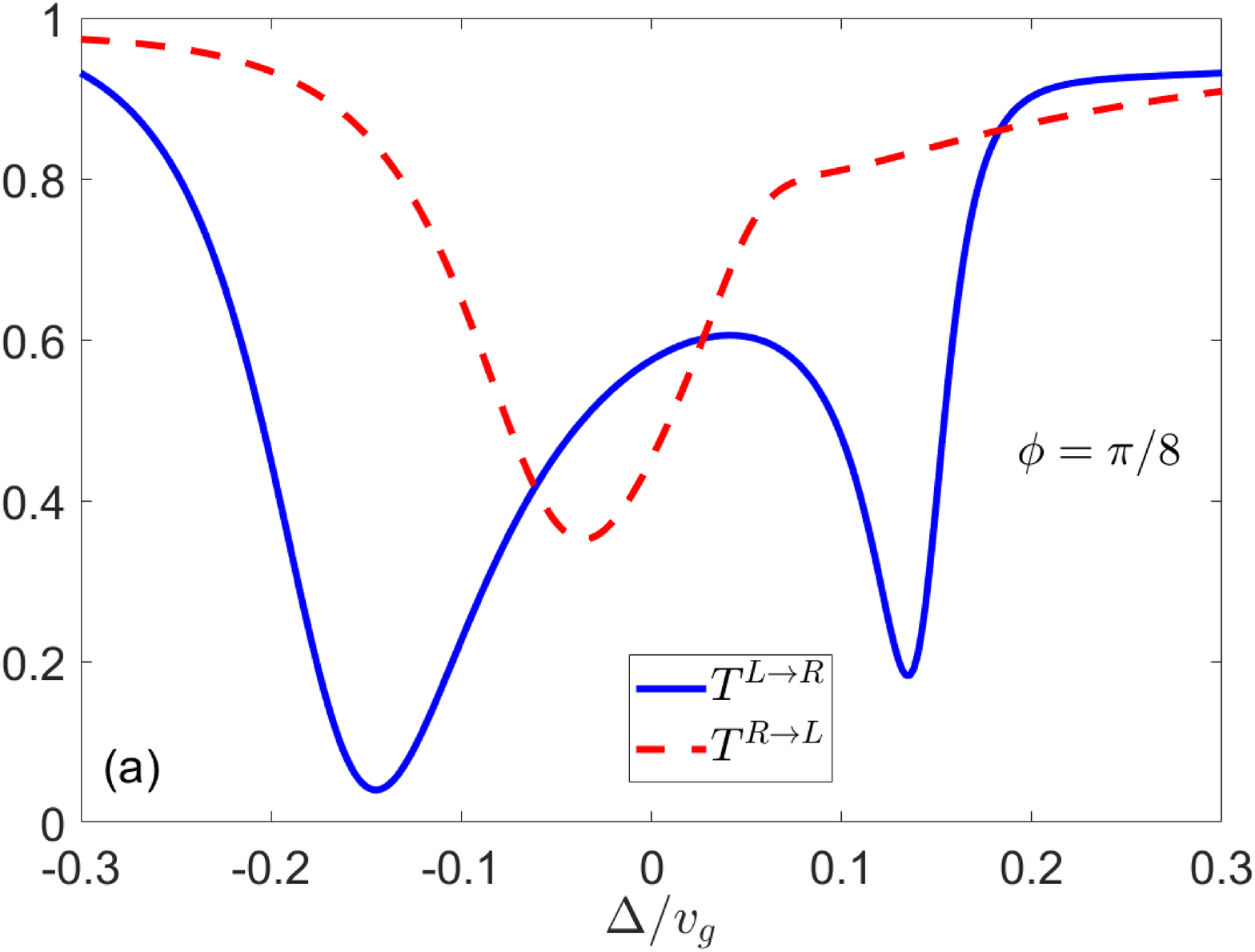}
\par\end{centering}
\begin{centering}
\includegraphics[width=1\columnwidth]{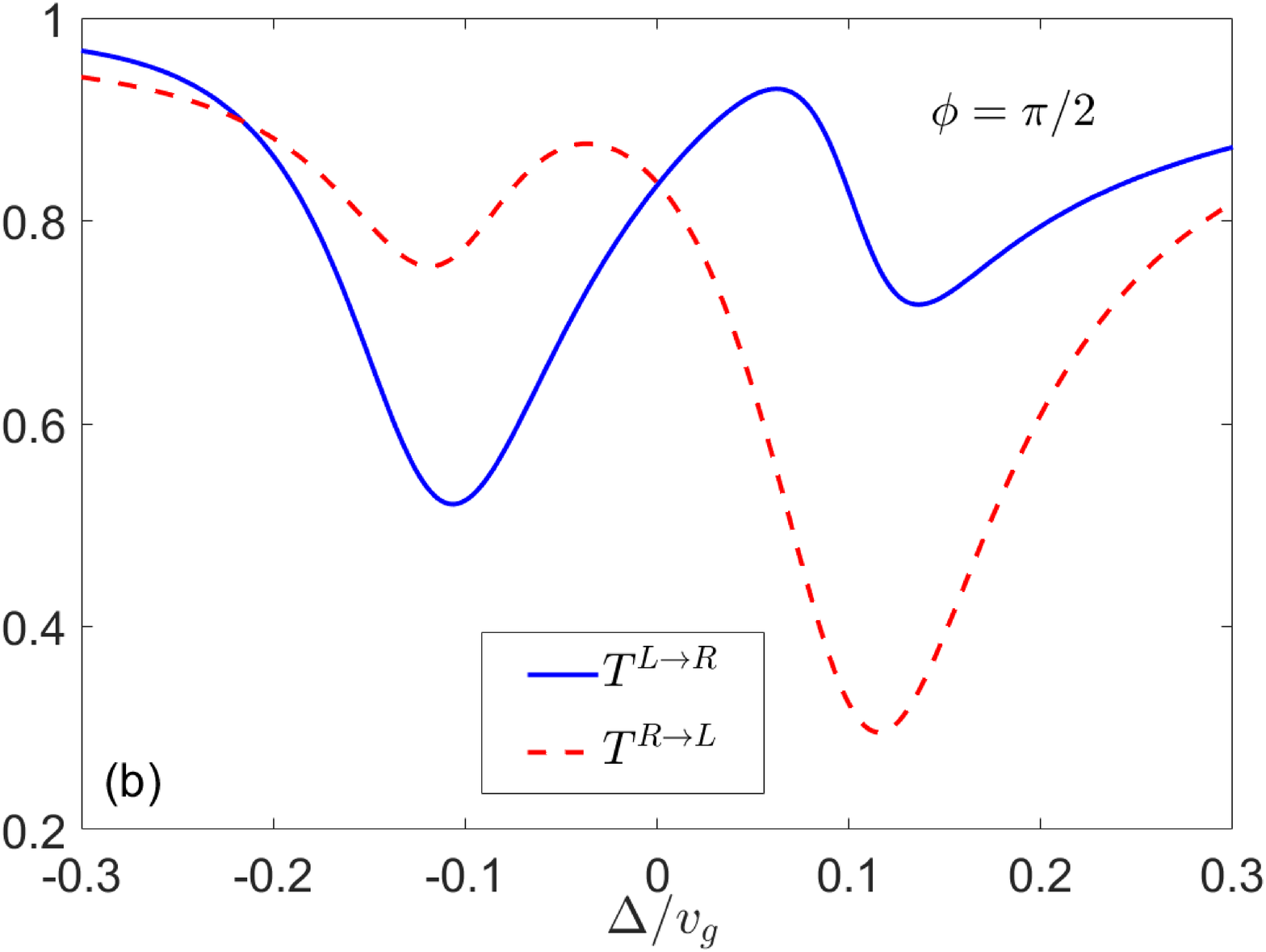}
\par\end{centering}
\caption{The transmission rates $T^{L\rightarrow R}$ and $T^{R\rightarrow L}$ as a function of the detuning $\Delta$ for (a) $\phi=\pi/8$ and (b) $\phi=\pi/2$. The other parameters are set as $(\lambda,f,g)=(0.1,0.3,0.2)v_g$ and $x_{0}=2$.}
\label{no0}
\end{figure}

\begin{figure}[tbp]
\begin{centering}
\includegraphics[width=1\columnwidth]{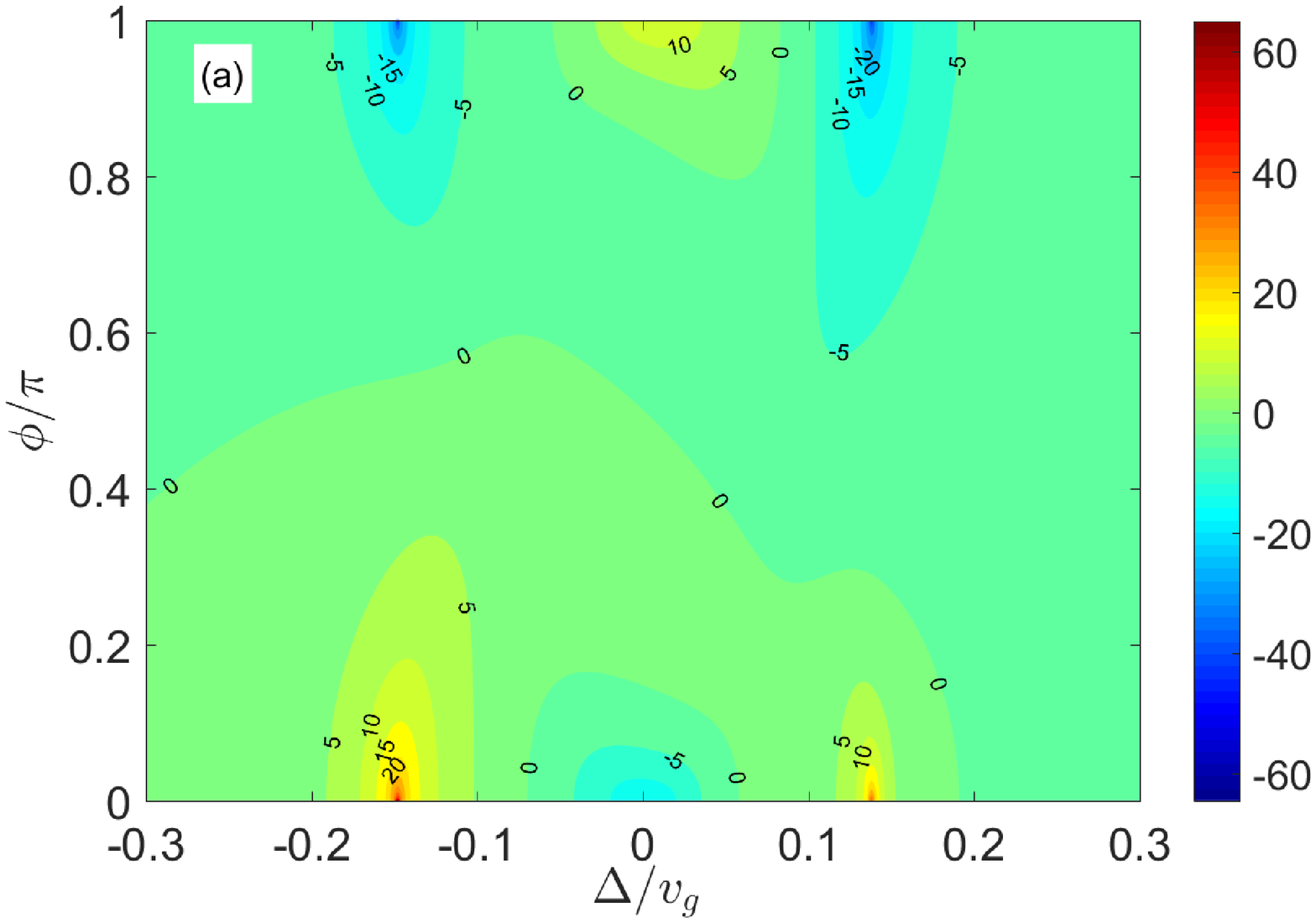}
\includegraphics[width=1\columnwidth]{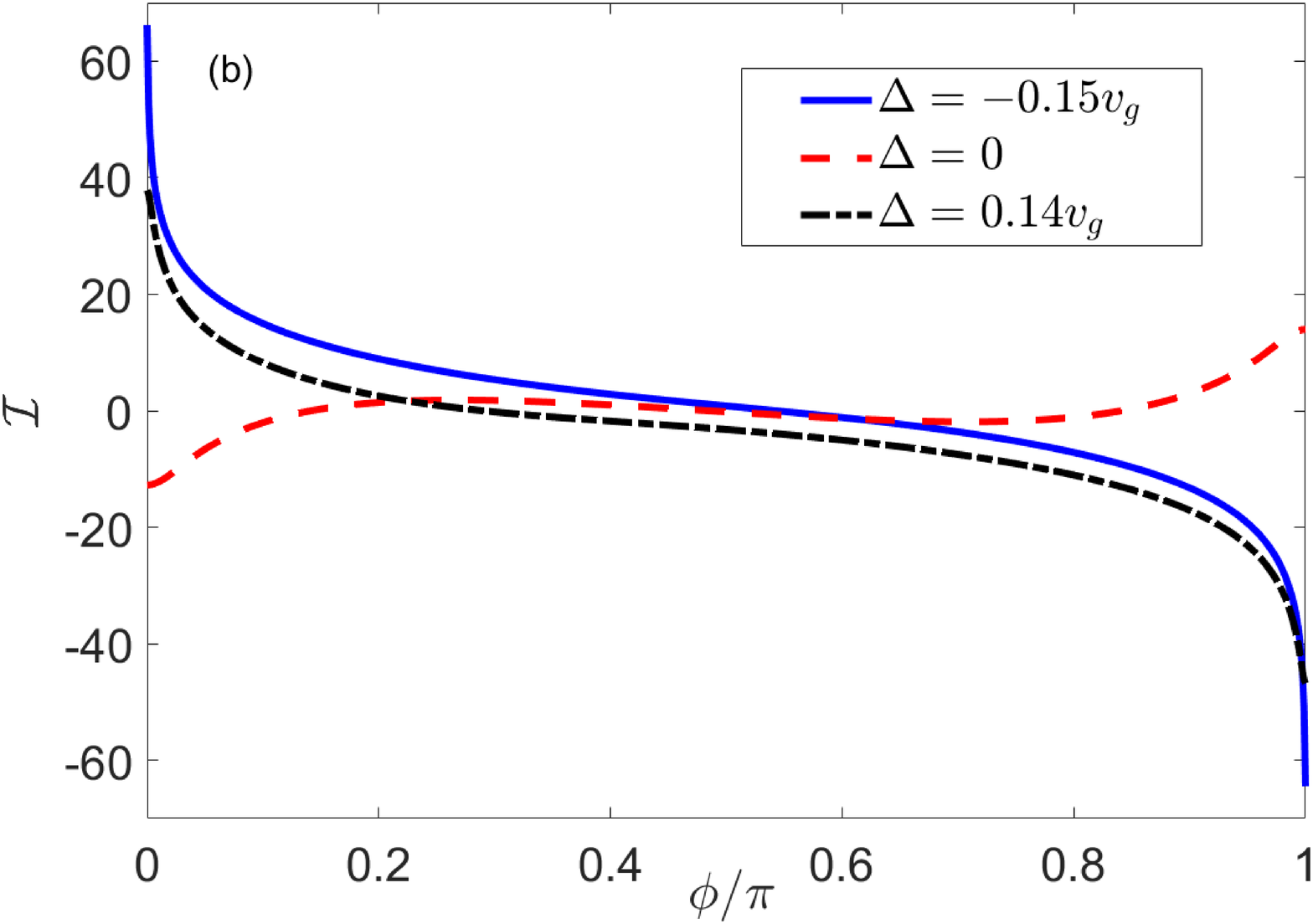}
\par\end{centering}
\caption{(a) The isolation ratio $\mathcal{I}$ as a function of the detuning $\Delta$ and phase $\phi$. {(b) The isolation ratio $\mathcal{I}$ as a function of the phase $\phi$ for different detunings.} The parameters are set as $(\lambda,f,g)=(0.1,0.3,0.2)v_g$ and $x_{0}=2$.}
\label{ratio}
\end{figure}

In Fig.~\ref{no0}, we plot the transmission rates $T^{L\rightarrow R}=|t|^{2}$ and
$T^{R\rightarrow L}=|t'|^{2}$ when the incident photons are
from the left and right sides as a function of the frequency of incident photons for different phase $\phi$ [the expressions of $t$ and $t'$ are given in Eqs.~(\ref{eq:tt}) and~(\ref{tp}) respectively]. It shows that
$T^{L\rightarrow R}$ can be either larger or smaller than $T^{R\rightarrow L}$, depending on the frequency of the incident photons and the phase. In other words, the phase can also be applied to adjust the single-photon nonreciprocal transmission behavior.

To quantitatively depict the non-reciprocality in our system, we define the isolation ratio $\mathcal{I}$ as
\begin{equation}
\mathcal{I}\,({\rm dB})=-10\times{\rm log}_{10}\frac{T^{L\rightarrow R}}{T^{R\rightarrow L}},
\end{equation}
and illustrate it as a function of the detuning $\Delta$ and the phase $\phi$ in Fig.~\ref{ratio}(a).  It shows that for the fixed detuning $\Delta=-0.15v_g$, the ratio $\mathcal{I}\approx20$\,dB can be achieved when the phase $\phi$ is tuned to be about $0.05\pi$. Correspondingly we achieve $T^{L\rightarrow R}/T^{R\rightarrow L}\approx0.01$. In other words, when the incident photons are from the right side, they will be transmitted to the left side with a relatively large probability, and will be nearly blocked when it is incident from the left side. However, for the same detuning $\Delta=-0.15v_g$, when the phase is tuned to $\phi\approx 0.95\pi$, we obtain $\mathcal{I}\approx-20$\,dB. That is, $T^{L\rightarrow R}/T^{R\rightarrow L}\approx100$, which implies that the left-going photons will be blocked. {We also demonstrate the phase modulation effect for different detunings in Fig.~\ref{ratio}(b). It shows that, for the  detunings $\Delta=-0.15v_g$ and $\Delta=0.14v_g$, the isolation rate can achieve as high as  $60\,(-60)$\,dB when the phase tends to be $0\,(\pi)$. As for the case of $\Delta=0$, this ratio can reach $\pm 20$\,dB.} Therefore, we have realized a controllable photon nonreciprocal transmission in our system where the phase plays as a sensitive controller.

{The underlying physics behind the single-photon nonreciprocal transmission can be revealed clearly in the momentum space.  For the sake of clarity, the expression of the Hamiltonian in the momentum space is derived in the Appendix. It is clear that when $i$ is replaced by $-i$, the Hamiltonian $\mathcal{H}$ in Eq.~(\ref{mh}) (in the Appendix) will be changed. This change cannot be deleted by any canonical transformation such as that given in the Appendix for $x_0\neq0$. That is, the emitter and the cavity locate at different points of the waveguide. In other words, the breaking of time-reversal symmetry leads to the single-photon nonreciprocal transmission.}

\section{Discussions}
\label{ds}

{The results in previous sections were restricted to the emitter-cavity resonant case, namely $\omega_a=\Omega=\omega$. It is worthwhile to study how the detuning between the emitter and the cavity affects those results. In addition, as shown in Eq.~(\ref{kx}) (in the Appendix), the factor $kx_0$ which disappears in the real space interaction Hamiltonian (\ref{VV}), plays as a coupling phase between the waveguide and the emitter. Therefore, the effect of different $x_0$ is also interesting. At last, a quantum system can never isolate from the surrounding environments. In our system, the interaction between the system and the bath will induce the decay of the cavity mode and the spontaneous emission of the emitter. In this section, we will demonstrate the effect of these factors to the photon transmission numerically.}

\begin{figure}[tbp]
\begin{centering}
\includegraphics[width=1\columnwidth]{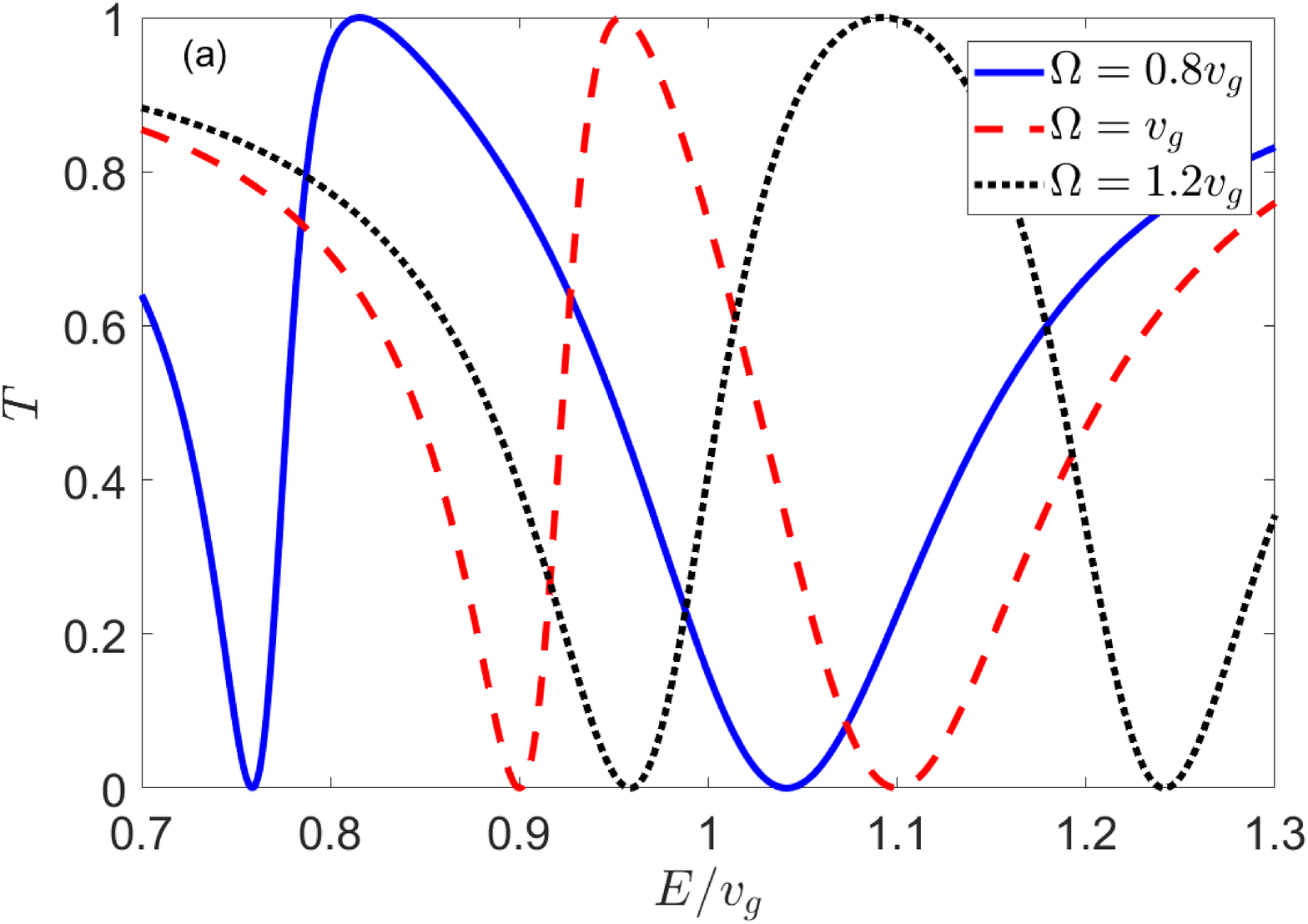}
\includegraphics[width=1\columnwidth]{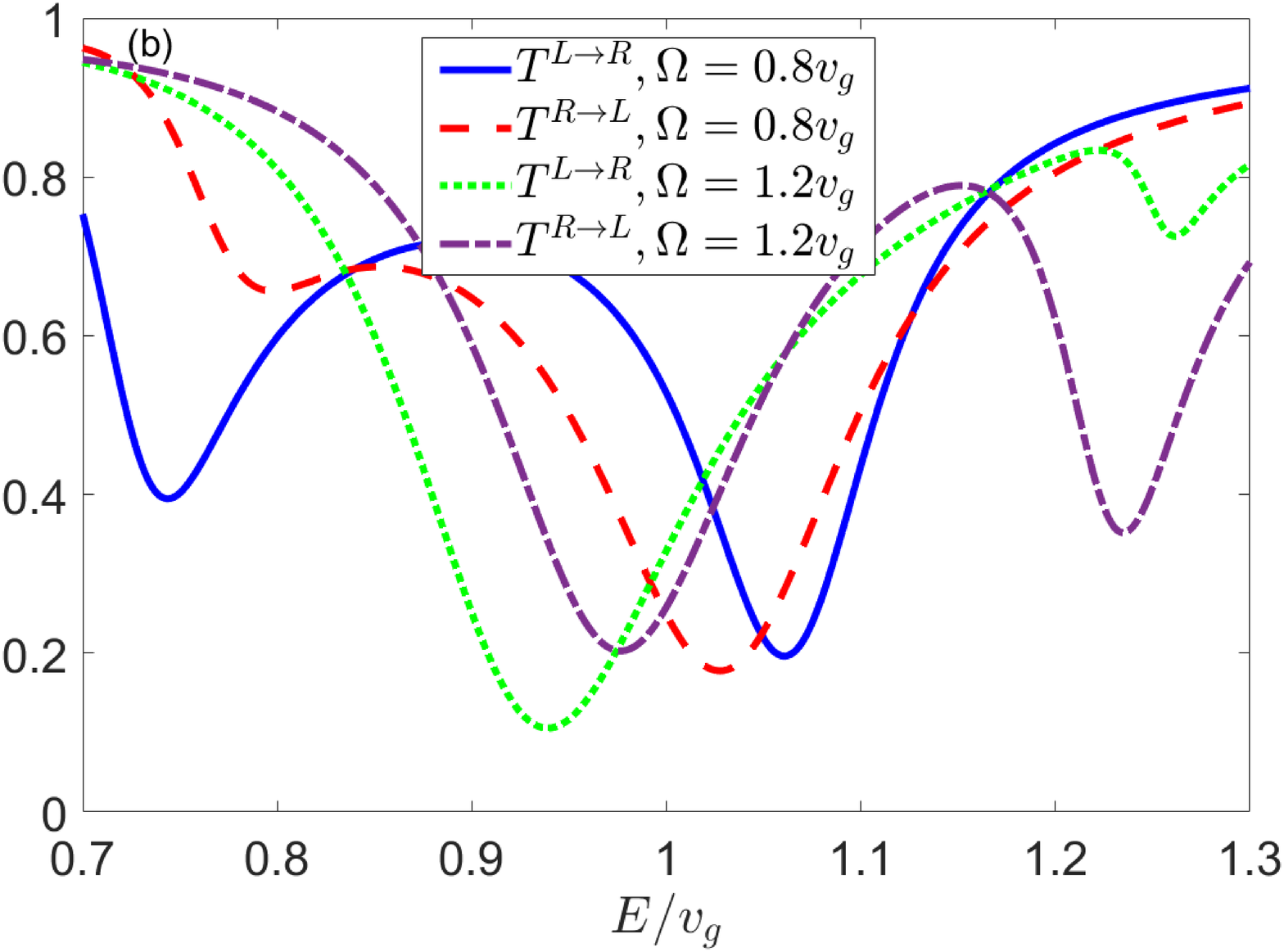}
\includegraphics[width=1\columnwidth]{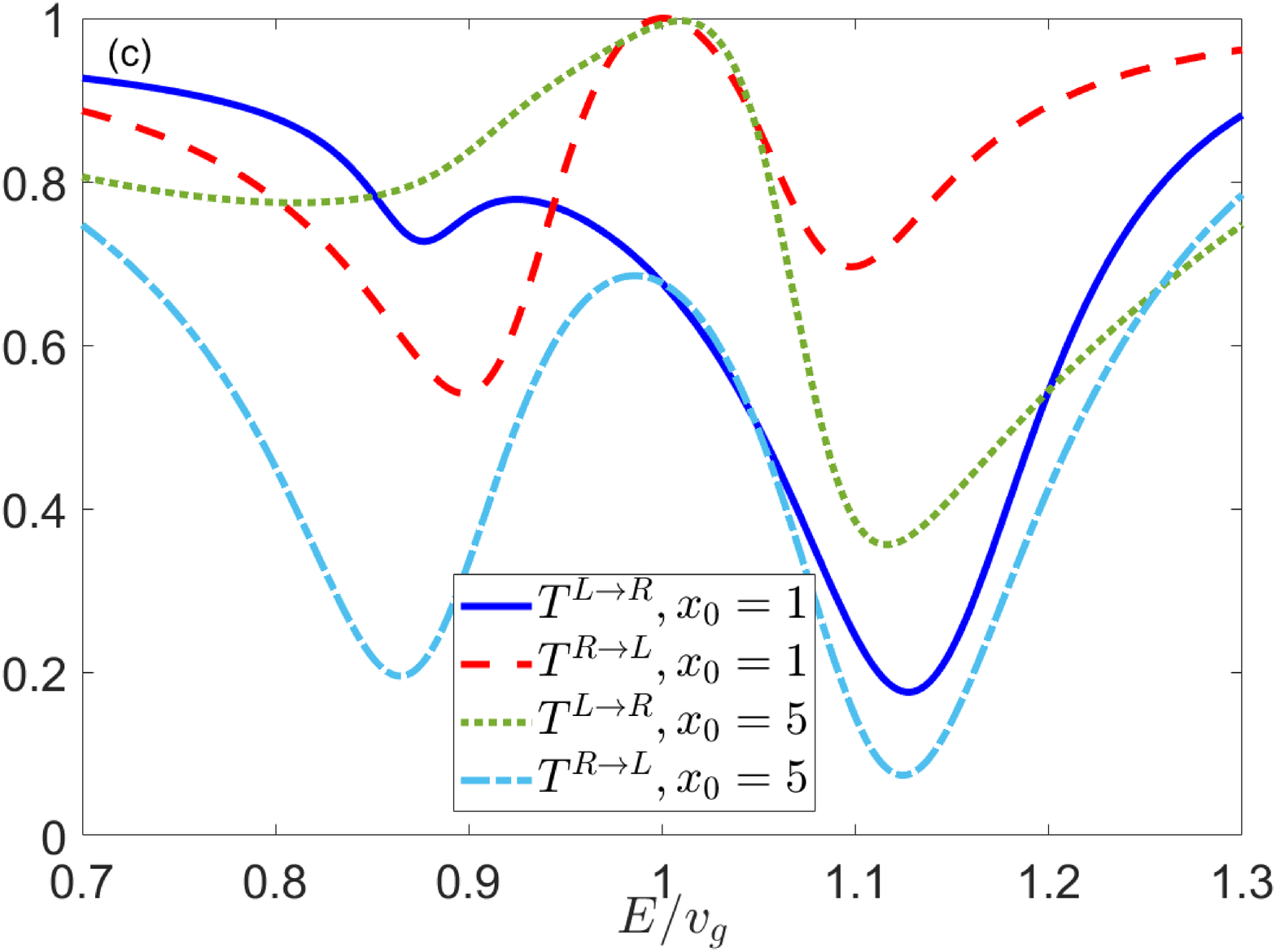}
\par\end{centering}
\caption{{The transmission rate $T$ as a function of the energy of the incident photon. The parameters are set as (a) $\omega_a=v_g, x_0=0$; (b) $\omega_a=v_g, x_0=2$;  (c) $\omega_a=\Omega=v_g$. The other parameters are $(\lambda,f,g)=(0.1,0.3,0.2)v_g$.}}
\label{dis1}
\end{figure}

{Now, let us consider the case with $\omega_a\neq\Omega$. In Fig.~\ref{dis1}(a), we plot the curves for the transmission rates with $x_0=0$. As discussed in the end of Sec.~\ref{one}, the time-reversal symmetry is not broken in this case, and we can not observe the non-reciprocal transmission. Due to the detuning between the emitter and the cavity,  we can observe that both the eigen-energies of the dressed states and the average energy of the quasidark state are shifted. The non-reciprocal transmission with $x_0\neq0$ is plotted in Fig.~\ref{dis1}(b), where we consider the cases with $\Omega$ being a little smaller or larger than $\omega_a$. It is shown that the broken time-reversal symmetry induced non-reciprocity is robust to the emitter-cavity detuning.}

{Furthermore, we also study the single-photon transmission behavior in Fig.~\ref{dis1}(c) for different $x_0$. It shows that the non-reciprocity is always present as long as $x_0\neq0$, that is, the emitter and the cavity locate at different points of the waveguide. The non-reciprocity is induced by the broken time-reversal symmetry and the isolation ratio depends not only on $x_0$ but also on the frequency of the incident photon.}

\begin{figure}[tbp]
\begin{centering}
\includegraphics[width=1\columnwidth]{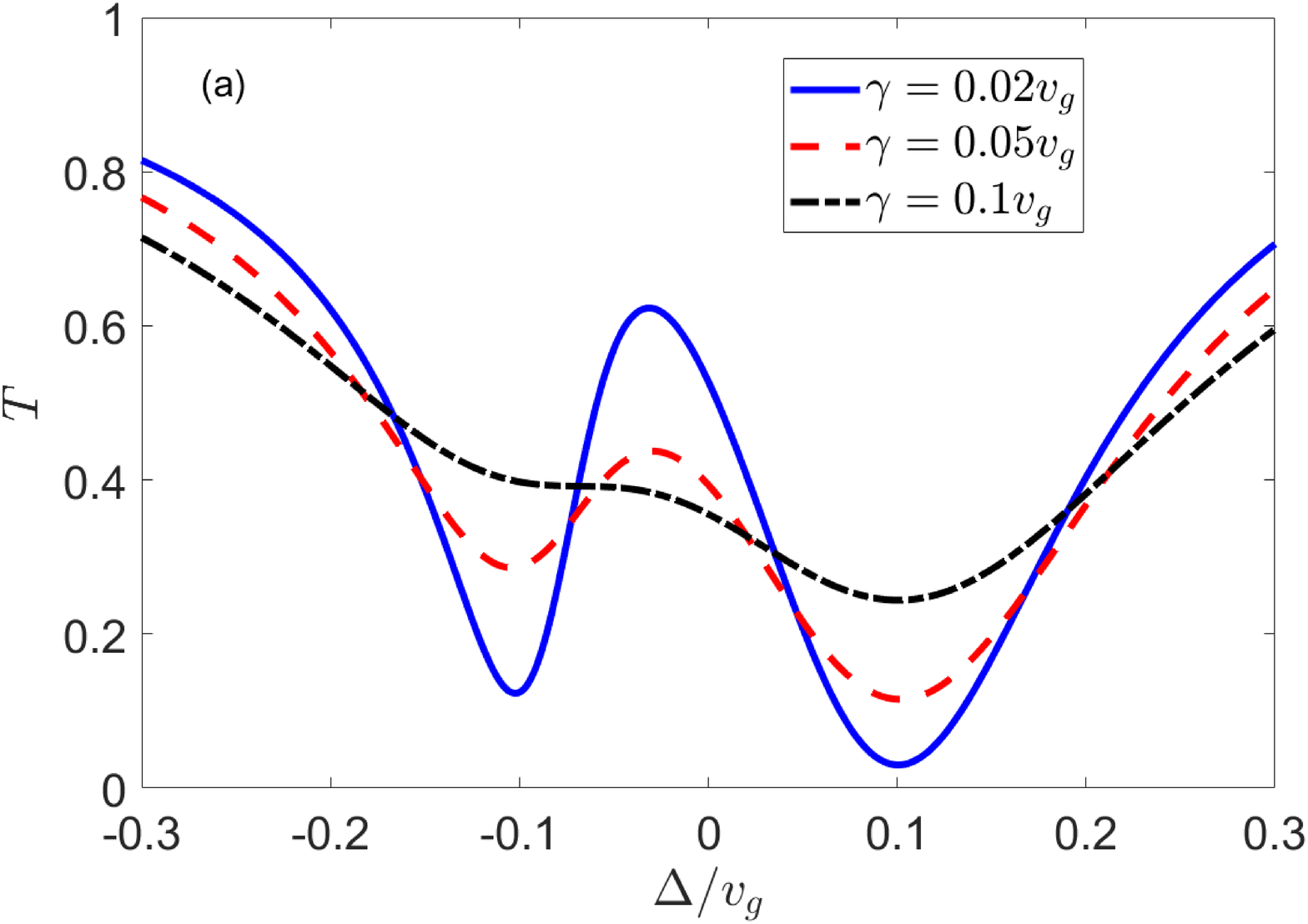}
\includegraphics[width=1\columnwidth]{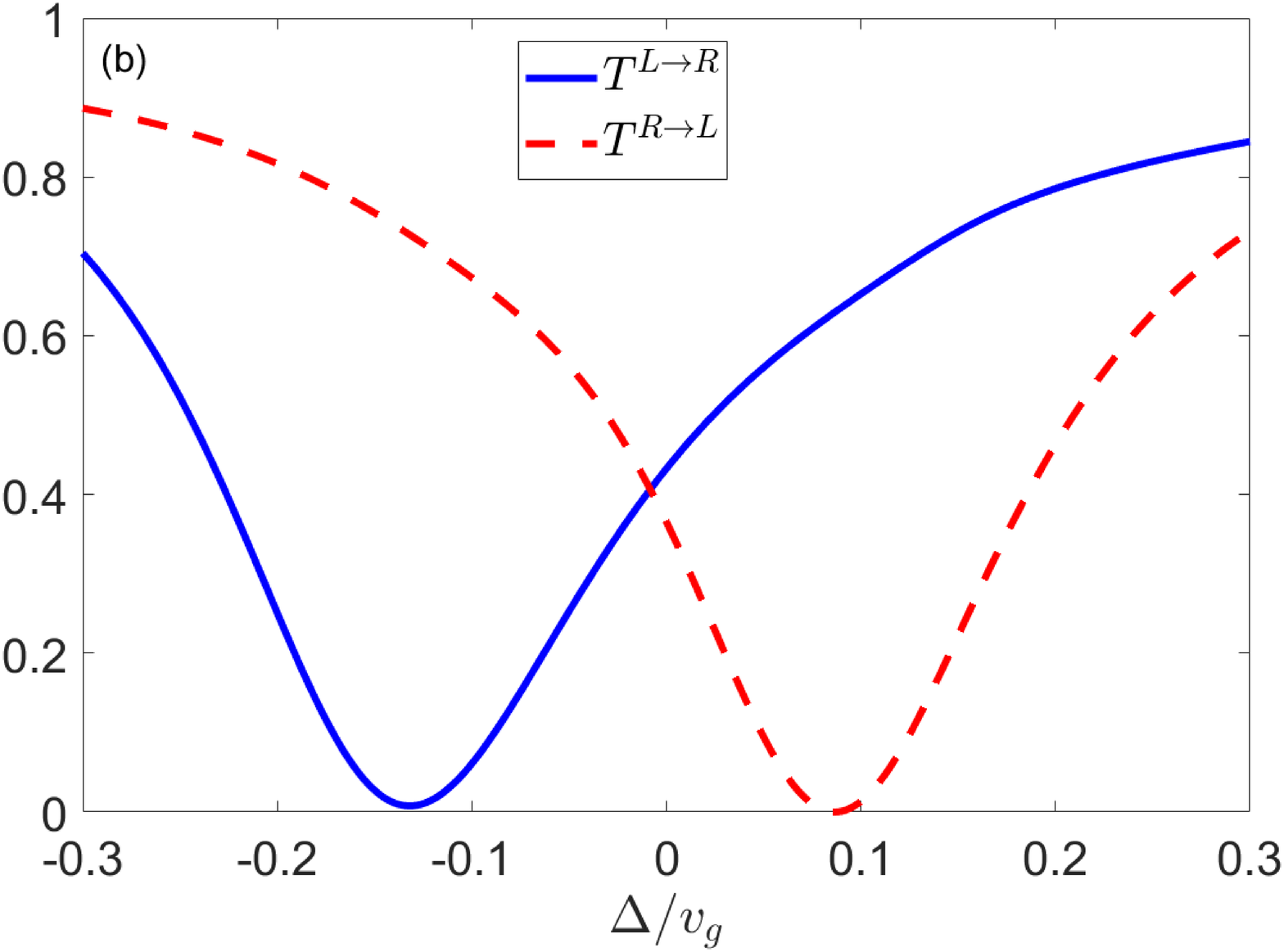}
\par\end{centering}
\caption{{The transmission rate $T$ as a function of $\Delta$. The parameters are set as (a) $\omega_a=\Omega=v_g, x_0=0$;  (b) $\omega_a=\Omega=v_g, \gamma=0.05v_g, x_0=2$. The other parameters are $(\lambda,f,g)=(0.1,0.3,0.2)v_g$.}}
\label{decay}
\end{figure}

{At last, we discuss the effect of the decay of the cavity mode and the spontaneous emission of the emitter on the photon transmission. For the sake of simplicity and to grasp the main physics, we will use the non-Hermitian Hamiltonian to describe the effect of the environments. That is, we will replace $\omega_a$ by $\omega_a-i\gamma_a$ and $\Omega$ by $\Omega-i\gamma_e$ respectively, where $\gamma_a$ is the decay rate of the cavity mode and $\gamma_e$ is the spontaneous emission rate of the emitter. In our discussions, we set $\gamma_a=\gamma_e=\gamma$. For $x_0=0$, the photon transmission is reciprocal, and we plot the transmission rates for different $\gamma$ by considering $\omega_a=\Omega=\omega$ in Fig.~\ref{decay}(a), where $\Delta=E-\omega$. For small $\gamma$ (for example $\gamma=0.02v_g$), we can also observe the Rabi splitting, in which the transmission rate will achieve its local minimum values at $\Delta=\pm\lambda$. By increasing $\gamma$, the valleys induced by the Rabi splitting become more and more flat, this is because the valleys are widened by the interaction between the system and the surrounding environments. Furthermore, we demonstrate the result for $x_0=2$ with a moderate $\gamma$ ($\gamma=0.05v_g$) in Fig.~\ref{decay}(b). Although the single photon scattering behavior has been modulated by the decay of the cavity and the spontaneous emission of the emitter, the non-reciprocity transmission is remain in that $T^{\rm{L\rightarrow R}}\neq T^{\rm{R\rightarrow L}}$ for $x_0\neq0$.}

\section{Experimental feasibility}
\label{proposal}
\begin{figure}[tbp]
\begin{centering}
\includegraphics[width=1\columnwidth]{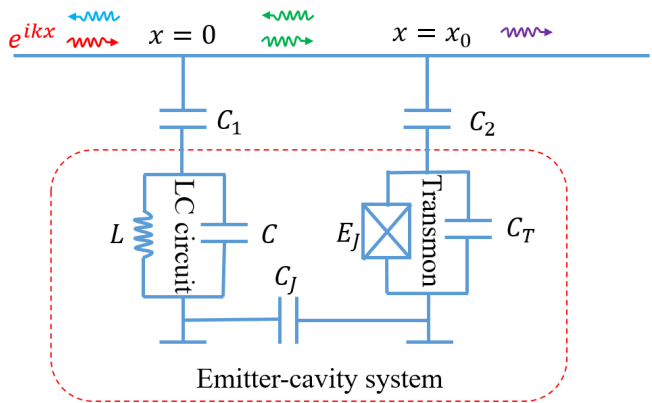}
\par\end{centering}
\caption{Schematic diagram for the possible system to implement single photon non-reciprocal transmission based on the circuit QED system: an LC circuit is coupled to a transmon qubit via the capacity $C_J$ and they are simultaneously coupled to a superconducting transmission line, which supports many bosonic modes with linear dispersion relation.}
\label{exp}
\end{figure}

To demonstrate our theoretical results about the non-reciprocal single-photon
transmission mechanism, we now propose an experimentally accessible device
by using superconducting quantum circuits, which is schematically shown in Fig.~\ref{exp}. As shown inside the dashed rectangle frame in Fig.~\ref{exp}, an LC circuit and a transmon qubit are coupled to each other via the capacity $C_J$, and serve as the emitter-cavity-coupled system in our theoretical proposal. The Lagrangian for the coupled system inside the dashed rectangle frame is expressed as
\begin{equation}
\mathcal{L}=\frac{C}{2}\dot{\psi}_L^2-\frac{1}{2L}{\psi}_L^2+\frac{C_T}{2}\dot{\psi}_T^2+
E_J\cos(\frac{\psi_T}{\phi_0})+\frac{C_J}{2}(\dot{\psi}_L-\dot{\psi}_T)^2.
\end{equation}
Here, $\psi_L$ and $\psi_T$ are respectively the phases of the LC circuit and the transmon. $C$ and $L$ are the capacitance and inductance of the LC circuit, $C_T$ is the capacitance of the transmon and $E_J$ is its Jospheson energy. $\phi_0=\hbar/(2e)$ is the reduced flux quantum. We introduce the conjugate node charge $Q_i=\partial\mathcal{L}/{\partial \dot{\psi}_i}\,(i=L,T)$ and follow the standard quantization procedure, the Hamiltonian can be obtained as
\begin{eqnarray}
H_{LT}&=&Q_L\dot{\psi}_L+Q_T\dot{\psi}_T-\mathcal{L}\nonumber\\
&=&\frac{Q_L^2}{2C_{L0}}+\frac{Q_T^2}{2C_{T0}}+\frac{\psi_L^2}{2L}-E_J\cos({\frac{\psi_T}{\phi_0}})
+\frac{Q_LQ_T}{C_{\rm int}},\nonumber\\ &&
\end{eqnarray}
where
\begin{eqnarray}
C_{L0}=\frac{(C_L+C_J)(C_T+C_J)-C_J^2}{C_T+C_J},\\
C_{T0}=\frac{(C_L+C_J)(C_T+C_J)-C_J^2}{C_L+C_J},\\
C_{\rm{int}}=\frac{(C_L+C_J)(C_T+C_J)-C_J^2}{C_J},
\end{eqnarray}
and it obeys the bosonic commutation relations $[\psi_m,Q_n]=i\delta_{m,n}\,(m,n=L,T)$. Furthermore, we denote $E_{TC}=e^2/(2C_{T0}),E_{LC}=e^2/(2C_{L0}),E_{LJ}=\phi_0^2/L$, and introduce the dimensionless charge and phase variables by
\begin{eqnarray}
\frac{Q_L}{2e}&=&(\frac{E_{LJ}}{8E_{LC}})^{\frac{1}{4}}q_L,\,\frac{\psi_L}{\phi_0}
=(\frac{8E_{LC}}{E_{LJ}})^{\frac{1}{4}}\varphi_L,\\
\frac{Q_T}{2e}&=&(\frac{E_J}{8E_{TC}})^{\frac{1}{4}}q_T,\,\frac{\psi_T}{\phi_0}
=(\frac{8E_{TC}}{E_{J}})^{\frac{1}{4}}\varphi_T.
\end{eqnarray}
The phase and charge operators can be furthermore expressed in the annihilation and creation operators, i.e., $\varphi_m=(a_m e^{-i\phi_m}+a_m^\dagger e^{i\phi_m})/\sqrt{2},
\,q_m=i(a_m^\dagger e^{i\phi_m}-a_m e^{-i\phi_m})/\sqrt{2}$ for $m=L,T$ and $\phi_T=0,\phi_L=\phi$. The Hamiltonian can then be approximately written as (in the situation of $E_J\gg E_{TC}$)
\begin{eqnarray}
H_{LT}&\approx&\hbar\omega_La_L^\dagger a_L+(\hbar\omega_T-E_{TC})a_T^\dagger a_T-\frac{E_{TC}}{2}a_T^{\dagger2} a_T^2\nonumber \\&&+J(a_T^\dagger a_L e^{-i\phi}+a_L^\dagger a_T e^{i\phi})+o[\epsilon^4],
\label{eH}
\end{eqnarray}
where we have applied the rotating wave approximation and define the parameters as $\omega_L=\sqrt{8E_{LC}E_{LJ}}/\hbar,\omega_T=\sqrt{8E_{TC}E_{J}}/\hbar,
\epsilon={8E_{TC}}/{E_{J}}$, and
\begin{equation}
J=\frac{2e^2}{C_{\rm int}}(\frac{E_{LJ}E_{J}}{64E_{LC}E_{TC}})^{\frac{1}{4}}.
\end{equation}

In our previous studies, we have assumed the resonant situation in which $\omega_L=\omega_T-E_{TC}/\hbar$ {(even in some parts of Sec.~\ref{ds}, we still consider a small detuning situation in which $|\omega_L-(\omega_T-E_{TC}/\hbar) |\ll E_{TC}/\hbar$)}. In this case, the self-nonlinear interaction, which is denoted by the last term in the first line of Hamiltonian~(\ref{eH}), induces a large detuning between the $|2\rangle_T\leftrightarrow|1\rangle_T$ (here, $|m\rangle_T$ represents the Fock state with $m$ photons for the transmon) transitions and LC circuit. Therefore, we can safely restrict ourselves in the vacuum and single photon subspace and therefore the transmon serves as a two-level emitter.  Then, the Hamiltonian becomes
\begin{eqnarray}
H_{LT}&\approx&\hbar\omega_La_L^\dagger a_L+(\hbar\omega_T-E_{TC})\sigma_z\nonumber \\&&+J(\sigma_+a_Le^{-i\phi}+h.c.),\nonumber\\
\label{eH1}
\end{eqnarray}
where $\sigma_z=|1\rangle_T {_T}\langle1|-|0\rangle_T {_T}\langle0|,
\sigma_+=|1\rangle_T {_T}\langle0|$.
At last, we introduce a superconducting transmission line, which supports multiple bosonic modes with linear dispersion relation. The transmission line is connected to the LC circuit and the transmon qubit simultaneously, via the capacities $C_1$ and $C_2$ at the point $x=0$ and $x=x_0$. As discussed before, we can choose either $x_0=0$ or $x_0\neq0$ in the experiments. The dynamical process of the system can be used to study the single-photon nonreciprocal transmission.

In the realistic experimental setup, the charge energy and the Josephson energy of the transmon qubit can achieve $E_{TC}/(2\pi\hbar)=200\sim500$ MHz and $E_J/(2\pi\hbar)=6\sim10$ GHz. The resonant frequencies of both of the transmon and the LC circuit are in the order of GHz~\cite{transmon,transmon1}, and their coupling strength can be tuned from tens of MHz to hundreds of MHz. In addition, the recent experimental progress has make it possible to coupling them to the superconducting transmission line with coupling strength of tens of MHz~\cite{peng,BRJ}.

\section{Conclusion}
\label{con}
In conclusion, we have investigated the single-photon scattering in a waveguide which is coupled to the emitter-cavity system. The cyclic energy-level diagram in our system makes it possible to adjust the scattering behavior by only tuning the phase of the coupling strength between the emitter and the cavity. Specially,  motivated by the experimental realization of coherent coupling between the transmon qubit and transmission line, we propose a nonreciprocal scattering scheme by coupling the emitter-cavity system to a waveguide, with the emitter and the cavity locating at different points of the waveguide. In such a system, the time-reversal symmetry is broken.

Moreover, the waveguide QED system has become more popular in the field of light-matter interaction~\cite{waveguide1,waveguide2}. We hope that our study on phase controlled photon transmission can provide a new way to realize photonic coherent control in waveguide systems and can be extended to study the two-photon scattering procss~\cite{EV}. We also believe our study will be applicable to quantum acoustics, where the size of the emitter can be comparable to the wavelength of the phonons. In this case, the time-reversal symmetry becomes easy to be broken.

\begin{acknowledgments}
 This work is supported by the the NSFC (under Grants No.~11875011, No. 11774024, No. 11534002, No. 11874037, No. U1930402, and U1730449) and Educational Commission of Jilin Province of China (Grant No. JJKH20190266KJ).
\end{acknowledgments}

\appendix%\appendixpage
\addcontentsline{toc}{section}{Appendices}\markboth{APPENDICES}{}
\begin{subappendices}
\section{The Hamiltonian in the momentum space}
{In the main text, we have studied the single-photon scattering process based on the Hamiltonian in the real space [e.g. Eqs.~(\ref{eq:hs},\ref{inH},\ref{VV})]. As we have stated in Sec.~\ref{ds}, not only $\phi$ but also $kx_0$, where $k$ is the wave vector, play roles as the effective coupling phases. To clarify this fact more clearly, we now derive the expression of the Hamiltonian in the momentum space. Following the similar approach used in Ref.~\cite{trans3}, we introduce the Fourier transformation
\begin{equation}
c_R(x)=\sum_{k_R}c_{k_R}e^{ik_Rx},\,\,c_L(x)=\sum_{k_L}c_{k_L}e^{ik_Lx},
\end{equation}
for the operators $c_R(x),\, c_L(x)$ and their Hermitian conjugates in Eqs.~(\ref{inH})
and ~(\ref{VV}). $c_{k_R}$ is the annihilation operator for the right-going photon with wave vector $k_{R}\,(>0)$, and $c_{k_L}$ is the annihilation operator for the left-going photon with wave vector $k_{L}\,(<0)$. In terms of $c_{k_R}$ and $c_{k_L}$, the Hamiltonian can be written as
\begin{equation}
\mathcal{H}=\mathcal{H}_s+\mathcal{H}_w+\mathcal{V},\label{mh}
\end{equation}
where
\begin{eqnarray}
\mathcal{H}_s&=&\omega_{a}a^{\dagger}a+\Omega|e\rangle\langle e|+\lambda(e^{i\phi}a^{\dagger}\sigma_{-}+h.c.),\\
\mathcal{H}_w&=&\sum_{k_R} \omega_{k_R}c_{k_R}^{\dagger} c_{k_R}+\sum_{k_L} \omega_{k_L}c_{k_L}^{\dagger} c_{k_L}, \\
\mathcal{V}&=&f\left(\sum_{k_R}a^{\dagger}c_{k_R}+\sum_{k_L}a^{\dagger}c_{k_L}+h.c.\right)\nonumber \\&+&g\left(\sum_{k_R}\sigma_+c_{k_R}e^{ik_Rx_0}
+\sum_{k_L}\sigma_+c_{k_L}e^{ik_Lx_0}+h.c.\right)\nonumber \\
\end{eqnarray}
with $\omega_{k_R}=v_gk_R$ and $\omega_{k_L}=-v_gk_L$. Therefore, $\phi$ plays a role as the cavity-emitter coupling phase and $kx_0$ plays the role
of emitter-waveguide coupling phase.}

{We now introduce the gauge to eliminate the phase in the emitter-resonator coupling $a\rightarrow ae^{i\phi}$, in this way, the Hamiltonian becomes $\mathcal{H}'_w=\mathcal{H}_w$ and
\begin{eqnarray}
\mathcal{H}'_s&=&\omega_{a}a^{\dagger}a+\Omega|e\rangle\langle e|+\lambda(a^{\dagger}\sigma_{-}+h.c.),\\
\mathcal{V}'&=&f\left(\sum_{k_R}a^{\dagger}c_{k_R}e^{-i\phi}+\sum_{k_L}a^{\dagger}
c_{k_L}e^{-i\phi}+h.c.\right)\nonumber\\
&&+g\left(\sum_{k_R}\sigma_+c_{k_R}e^{ik_Rx_0}
+\sum_{k_L}\sigma_+c_{k_L}e^{ik_Lx_0}+h.c.\right).\label{kx}\nonumber \\
\end{eqnarray}
It is obvious that $\phi$ now plays a role as the cavity-waveguide coupling phase
and $kx_0$ plays the role of emitter-waveguide coupling phase.}

{Alternatively, we can also perform the gauge $\sigma_-\rightarrow\sigma_-e^{i\phi}$, then the Hamiltonian become $\mathcal{H}''_s=\mathcal{H}'_s$ and
\begin{eqnarray}
\mathcal{V}''&=&f\left(\sum_{k_R}a^{\dagger}c_{k_R}+\sum_{k_L}a^{\dagger}c_{k_L}+h.c.
\right)\nonumber\\&&
+g\left(\sum_{k_R}\sigma_+c_{k_R}e^{i(k_Rx_0+\phi)}
+\sum_{k_L}\sigma_+c_{k_L}e^{i(k_Lx_0+\phi)}+h.c.\right),\nonumber \\
\end{eqnarray}
and $k_{(L/R)}x_0+\phi$ serves as the phase of emitter-waveguide coupling.}

\end{subappendices}


\begin{thebibliography}{99}
\bibitem{opto1} M. Hafezi and P. Rabl, \textit{Optomechanically induced non-reciprocity in microring resonators},  Opt. Express {\bf 20}, 7672 (2012).

\bibitem{opto2} Z. Shen, Y. L. Zhang, Y. Chen, C. L. Zou, Y. F. Xiao, X. B.
Zou, F. W. Sun, G. C. Guo, and C. H. Dong, \textit{Experimental realization of optomechanically induced non-reciprocity}, Nature
Photon. {\bf 10}, 657 (2016).

\bibitem{opto3}S. Manipatruni, J. T. Robinson, and M. Lipson,  \textit{Optical nonreciprocity in optomechanical structures}, Phys. Rev. Lett. {\bf 102},
213903 (2009).

\bibitem{opto4}Z. Wang, L. Shi, Y. Liu, X. Xu, and X. Zhang,
\textit{Optical nonreciprocity in asymmetric optomechanical couplers}, Sci. Rep. {\bf 5}, 8657 (2015).

\bibitem{opto5}E. Verhagen and A. Al$\rm\grave{u}$, \textit{Optomechanical nonreciprocity},  Nat. Phys. {\bf 13}, 922 (2017).

\bibitem{opto6}X. W. Xu and Y. Li, \textit{Optical nonreciprocity and optomechanical circulator in three-mode optomechanical systems}, Phys. Rev. A {\bf 91}, 053854 (2015).

\bibitem{opto7}X. W. Xu, Y. Li, A. X. Chen, and Y. X. Liu, \textit{Nonreciprocal conversion between microwave and optical photons in electro-optomechanical systems}, Phys. Rev. A {\bf 93}, 023827 (2016).

\bibitem{opto8}G. Li, X. Xiao, Y. Li, and X. G. Wang,  \textit{Tunable optical nonreciprocity and a phonon-photon router in an optomechanical system with coupled mechanical and optical modes}, Phys. Rev. A {\bf 97}, 023801 (2018).

\bibitem{opto9}C. Jiang, L. N. Song, and Y. Li, \textit{Directional amplifier in an optomechanical system with optical gain}, Phys. Rev. A {\bf 97}, 053812 (2018).

\bibitem{opto10} K. Fang, J. Luo, A. Metelmann, M. H. Matheny, F. Marquardt,
A. A. Clerk, and O. Painter, \textit{Generalized non-reciprocity in an optomechanical circuit via synthetic magnetism and reservoir engineering}, Nature Phys. {\bf 13}, 465 (2017).

\bibitem{opto11}{S. Barzanjeh, M. Wulf, M. Peruzzo, M. Kalaee, P. B. Dieterle, O. Painter, and J. M. Fink, \textit{Mechanical on-chip microwave circulator}, Nat. Commun. {\bf 8}, 953 (2017).}

\bibitem{opto12}{H. Xu, D. Mason, L. Jiang, and J. G. E. Harris, \textit{Topological energy transfer in an optomechanical system with exceptional points}, Nature {\bf 537}, 81 (2016).}


\bibitem{PT1} N. Bender, S. Factor, J. D. Bodyfelt, H. Ramezani, D. N.
Christodoulides, F. M. Ellis, and T. Kottos, \textit{Observation of asymmetric transport in structures with active nonlinearities}, Phys. Rev. Lett. {\bf 110}, 234101 (2013).
\bibitem{PT2}B. Peng, S. K. $\rm\ddot{O}$zdemir, F. Lei, F. Monifi, M. Gianfreda,
G. L. Long, S. Fan, F. Nori, C. M. Bender, and L. Yang, \textit{Parity-time-symmetric whispering-gallery microcavities},  Nat. Phys. {\bf 10}, 394 (2014).

\bibitem{PT3} L. Chang, X. Jiang, S. Hua, C. Yang, J. Wen, L.
Jiang, G. Li, G. Wang, and M. Xiao, \textit{Parity-time symmetry and variable optical isolation in active-passive-coupled microresonators}, Nat. Photonics {\bf 8}, 524 (2014).

\bibitem{PT4}K. G. Makris, R. E.-Ganainy, D. N. Christodoulides, and Z. H. Mussliman, \textit{$\mathcal{PT}$-symmetric optical lattices}, Phys. Rev. A {\bf 81}, 063807(2010).

\bibitem{PT5}C. E. R$\rm \ddot{u}$ter, K. G. Makris, R. E.-Ganainy, D. N. Christodoulides, M. Segev, and  D. Kip, \textit{Observation of parity-time symmetry in optics}, Nature Phys. {\bf 6}, 192 (2010).

\bibitem{PT6}L. F. Xue, Z. R. Gong, H. B. Zhu, and Z. H. Wang, \textit{$\mathcal{PT}$ symmetric phase transition and photonic transmission in an optical trimer system}, Opt. Express {\bf 25}, 17249 (2017).

\bibitem{nonl1}E. Mascarenhas, D. Gerace, D. Valente, S. Montangero, A. Auff$\rm\acute{e}$ves, and M. F. Santos, \textit{A quantum optical valve in a nonlinear-linear resonators junction}, Europhys. Lett. {\bf 106}, 54003 (2014).

\bibitem{nonl2} M. Krause, H. Renner, and E. Brinkmeyer, \textit{Optical isolation in silicon waveguides based on nonreciprocal Raman amplification},  Electron. Lett. {\bf 44}, 691 (2008).

\bibitem{nonl3} M. D. Tocci, M. J. Bloemer, M. Scalora, J. P. Dowling, C. M. Bowden, \textit{Thin-film nonlinear optical diode}, Appl. Phys. Lett. {\bf 66},  2324 (1995).

\bibitem{nonl4} L. Fan, J. Wang, L. T. Varghese, H. Shen, B. Niu, Y Xuan, A. M. Weiner, and  M. Qi, \textit{An all-silicon passive optical diode}, Science {\bf 335}, 447 (2012).

\bibitem{nonl5} Y. Shi, Z. F. Yu, and S. H. Fan, \textit{Limitations of nonlinear optical isolators due to dynamic reciprocity},  Nature Photon. {\bf 9}, 388 (2015).

\bibitem{nonl6} J. Zhang, B. Peng, S. K. $\rm{\ddot{O}}$zdemir, Y. X. Liu, H. Jing, X. Y.  L$\rm{\ddot{u}}$, Y. L.  Liu, L. Yang, and F. Nori, \textit{Giant nonlinearity via breaking parity-time symmetry: A route to low-threshold phonon diodes},
Phys. Rev. B {\bf 92}, 115407 (2015).

\bibitem{nonl7} L. Fan, L. T. Varghese, and J. Wang, \textit{Silicon optical diode with 40 dB nonreciprocal transmission}, Opt. Lett. {\bf 38},  1259 (2013).

\bibitem{nonl8}H. Z. Shen, Y. H. Zhou, X. X. Yi, \textit{Tunable photon blockade in coupled semiconductor cavities}, Phys. Rev. A {\bf 91}, 063808 (2015).

\bibitem{nonl9}A. S. Zheng, G. Y.  Zhang, H. Y. Chen, T. T. Mei, and J. B. Liu, \textit{Nonreciprocal light propagation in coupled microcavities system beyond weak-excitation approximation}, Sci. Rep. {\bf 7}, 14001 (2017) .

\bibitem{nonl10}L. N. Song, Z. H. Wang, and  Y. Li, \textit{Enhancing optical nonreciprocity by an atomic ensemble in two coupled cavities},  Optics Communications {\bf 415}, 39 (2018).

\bibitem{nonl11}R. Huang,  A. Miranowicz,  J.-Q. Liao, F. Nori, and H. Jing,
 \textit{Nonreciprocal photon blockade}, Phys. Rev. Lett. {\bf 121}, 153601 (2018).

\bibitem{nonl12}L. Tang, J. Tang, W. Zhang, G. Lu, H. Zhang, Y. Zhang, K. Xia, and M. Xiao, \textit{On-Chip chiral single-photon interface: isolation and unidirectional emission}, Phys. Rev. A {\bf 99}, 043833 (2019).

\bibitem{nonl13}{J. Dai, A. Roulet, H. N. Le, and V. Scarani, \textit{Rectification of light in the quantum regime}, Phys. Rev. A {\bf 92}, 063848 (2015).}

\bibitem{atomic1} D. W. Wang, H. T. Zhou, M. J. Guo, J. X. Zhang, J. Evers, and S. Y. Zhu, \textit{Optical diode made from a moving photonic crystal}, Phys. Rev. Lett. {\bf 110}, 093901 (2013).

\bibitem{atomic2}J.-H. Wu, M. Artoni, and G. C. La Rocca, \textit{Non-Hermitian degeneracies and unidirectional reflectionless atomic lattices}, Phys. Rev. Lett. {\bf 113}, 123004 (2014).

\bibitem{atomic3}X.-D. Bai, B. A. Malomed, and F.-G. Deng, \textit{Unidirectional transport of wave packets through tilted discrete breathers in nonlinear lattices with asymmetric defects},  Phys. Rev. E {\bf 94}, 032216 (2016).

\bibitem{atomic4}Y. Huang, Y. Shen, C. Min, S. Fan, and G. Veronis, \textit{Unidirectional reflectionless light propagation at exceptional points}, Nanophotonics {\bf 6}, 97 (2017).

\bibitem{atomic5}Y.-M. Liu, F. Gao, C.-H. Fan, and J.-H. Wu, \textit{Asymmetric light diffraction of an atomic grating with PT symmetry}, Opt. Lett. {\bf 42}, 4283 (2017).

\bibitem{atomic6}S. Zhang, Y. Hu, G. Lin, Y. Niu, K. Xia, J. Gong, and  S. Gong, \textit{Thermal-motion-induced non-reciprocal quantum optical system},
Nature Photon. {\bf 12}, 744 (2018).

\bibitem{trans1}J. T. Shen and S. Fan, \textit{Coherent photon transport from spontaneous emission in one-dimensional waveguides}, Opt. Lett. {\bf 30}, 2001 (2005).

\bibitem{trans2}J. T. Shen and S. Fan, \textit{Coherent single photon transport in a one-dimensional waveguide coupled with superconducting quantum bits}, Phys. Rev. Lett. {\bf 95}, 213001 (2005).

\bibitem{trans3}J. T. Shen and S. Fan, \textit{Theory of single-photon transport in a single-mode waveguide. I. Coupling to a cavity containing a two-level atom}, Phys. Rev. A {\bf 79}, 023837 (2009).

\bibitem{trans4}L. Zhou, Z. R. Gong, Y. X. Liu, C. P. Sun, and F. Nori, \textit{Controllable scattering of a single photon inside a one-dimensional resonator waveguide}, Phys. Rev. Lett. {\bf 101}, 100501 (2008).

\bibitem{trans5}P. Longo, P. Schmitteckert, and K. Busch, \textit{Few-photon transport in low-dimensional systems: interaction-induced radiation trapping}, Phys. Rev. Lett. {\bf 104}, 023602 (2010).

\bibitem{trans6}Z. H. Wang, Y. Li, D. L. Zhou, C. P. Sun, and P. Zhang, \textit{Single-photon scattering on a strongly dressed atom}, Phys. Rev. A {\bf 86}, 023824 (2012).

\bibitem{trans7}L. Zhou, L. P. Yang, Y. Li, and C. P. Sun, \textit{Quantum routing of single photons with a cyclic three-level system}, Phys. Rev. Lett. {\bf 111},
103604 (2013).

\bibitem{trans8} T. Tian, D. Z. Xu, T. Y. Zheng, and C. P. Sun, \textit{Coherent control of single photons in the cross resonator arrays via the dark state mechanism},  Eur. Phys. J. D {\bf 67}, 69 (2013).

\bibitem{trans9}Y. Ma, T. Li, and A. Zheng, \textit{Single-photon frequency conversion for generation of three-dimensional entanglement}, Laser Phys. {\bf 28},  075202 (2018).

\bibitem{trans10}W. Z. Jia, Y. W. Wang, and Y.-x. Liu, \textit{Efficient single-photon frequency conversion in the microwave domain using superconducting quantum circuits}, Phys. Rev. A {\bf 96}, 053832 (2017).

\bibitem{trans11}Z. H. Wang, L. Zhou, Y. Li, and C. P. Sun, \textit{Controllable single-photon frequency converter via a one-dimensional waveguide}, Phys. Rev. A {\bf 89}, 053813 (2014).

\bibitem{trans12}W.-B. Yan, J.-F. Huang, and H. Fan, \textit{Tunable single-photon frequency conversion in a Sagnac interferometer}, Sci. Rep. {\bf 3}, 3555 (2013).

\bibitem{gaint1} M. V. Gustafsson, T. Aref, A. F. Kockum, M. Ekstr$\rm{\ddot{o}}$m,
G. Johansson, and P. Delsing, \textit{Propagating phonons coupled to an artificial atom}, Science {\bf 346}, 207 (2014).

\bibitem{gaint2} M. V. Gustafsson, P. V. Santos, G. Johansson, and P. Delsing, \textit{Local probing of propagating acoustic waves in a gigahertz echo chamber},
Nat. Phys. {\bf 8}, 338 (2012).

\bibitem{gaint3} J. Koch, T. M. Yu, J. Gambetta, A. A. Houck, D. I. Schuster,
J. Majer, A. Blais, M. H. Devoret, S. M. Girvin, and R. J.
Schoelkopf, \textit{Charge-insensitive qubit design derived from the Cooper pair box}, Phys. Rev. A {\bf 76}, 042319 (2007).

\bibitem{gaint4} S. Datta, \textit{Surface Acoustic Wave Devices} (Prentice-Hall,
Englewood Cliffs, NJ, 1986).

\bibitem{gaint5}D. Morgan, \textit{Surface Acoustic Wave Filters}, 2nd ed. (Academic,
Amsterdam, 2007).

\bibitem{gaint6}A. F. Kockum, P. Delsing, and G. Johansson, \textit{Designing frequency-dependent relaxation rates and Lamb shifts for a giant artificial atom}, Phys. Rev. A {\bf 90}, 013837 (2014).

\bibitem{gaint7}L. Guo,  A. Grimsmo, A. F. Kockum, M. Pletyukhov, and G. Johansson, \textit{Giant acoustic atom: A single quantum system with a deterministic time delay},  Phys. Rev. A {\bf 95}, 053821 (2017).

\bibitem{giant8} A. F. Kockum, G. Johansson, and F. Nori, \textit{Decoherence-free interaction between giant atoms in waveguide quantum electrodynamics}, Phys. Rev. Lett. {\bf 120}, 140404 (2018).

\bibitem{JC}E. T. Jaynes and F. W. Cummings, \textit{Comparison of quantum and semiclassical radiation theories with application to the beam maser}, Proc. IEEE {\bf 51}, 89 (1963).

\bibitem{ZW}Z. Wang and D. L. Zhou, \textit{Quasidark state and quantum interference in the Jaynes-Cummings model with a common bath}, Phys. Rev. A {\bf 89}, 013809 (2014).
\bibitem{Tian} T. Tian, and L. J. Song, \textit{Collective effect of single-photon scattering on the three-level atomic array}, Europhys. Lett. {\bf  124}, 34002 (2018).

\bibitem{cy1} Y.-x. Liu, J. Q. You, L. F. Wei, C. P. Sun, and F. Nori, \textit{Optical selection rules and phase-dependent adiabatic state control in a superconducting quantum circuit}, Phys. Rev. Lett. {\bf 95}, 087001 (2005).

\bibitem{cy2} W. Z. Jia and L. F. Wei, \textit{Gains without inversion in quantum systems with broken parities}, Phys. Rev. A {\bf 82}, 013808 (2010).

\bibitem{cy3} M. Shapiro, E. Frishman, and P. Brumer, \textit{Coherently controlled asymmetric synthesis with a chiral light}, Phys. Rev. Lett. {\bf 84}, 1669 (2000).

\bibitem{cy4} P. Kr$\rm{\acute{a}}l$ and M. Shapiro, \textit{Cyclic population transfer in quantum systems with broken symmetry}, Phys. Rev. Lett. {\bf 87}, 183002 (2001).

\bibitem{cy5} P. Kr$\rm{\acute{a}}l$, I. Thanopulos, M. Shapiro, and D. Cohen, \textit{Two-step enantio-selective optical switch}, Phys. Rev. Lett. {\bf 90}, 033001 (2003).

\bibitem{cy6} Y. Li, C. Bruder, and C. P. Sun, \textit{Generalized Stern-Gerlach effect for chiral molecules}, Phys. Rev. Lett. {\bf 99}, 130403 (2007).

\bibitem{cy7}  J. Tang, W. Geng, and X. Xu, \textit{Quantum interference induced photon blockade in a coupled single quantum dot-cavity system},  Sci. Rep. {\bf 5}, 9252 (2015).

\bibitem{cy8}  W. Z. Jia, L. F. Wei, Y. Li, and Y.-x. Liu, \textit{Phase-dependent optical response properties in an optomechanical system by coherently driving the mechanical resonator}, Phys. Rev. A {\bf 91}, 043843 (2015).

\bibitem{cy9} X. W. Xu and Y. J. Li, \textit{Antibunching photons in a cavity coupled to an optomechanical system}, J. Phys. B {\bf 46}, 035502 (2013).


\bibitem{clerk}A. Metelmann and A. A. Clerk, \textit{Nonreciprocal photon transmission and amplification via reservoir engineering}, Phys. Rev. X {\bf 5}, 021025 (2015).

\bibitem{transmon}J. Koch, T. M. Yu, J. Gambetta, A. A. Houck, D. I. Schuster, J. Majer, A. Blais, M. H. Devoret, S. M. Girvin, and R. J. Schoelkopf, \textit{Charge-insensitive qubit design derived from the Cooper pair box}, Phys. Rev. A {\bf 76}, 042319 (2007).

\bibitem{transmon1}J. Majer, J. M. Chow, J. M. Gambetta, J. Koch, B. R. Johnson, J. A. Schreier, L. Frunzio, D. I. Schuster, A. A. Houck, A. Wallraff, A. Blais, M. H. Devoret, S. M. Girvin, and R. J. Schoelkopf, \textit{Coupling superconducting qubits via a cavity bus}, Nature {\bf 449},  443 (2007).

\bibitem{peng}Z. H. Peng, J. H. Ding, Y. Zhou, L. L. Ying, Z. Wang, L. Zhou, L. M. Kuang, Y.-x. Liu, O. V. Astafiev, and J. S. Tsai, \textit{Vacuum-induced Autler-Townes splitting in a superconducting artificial atom}, Phys. Rev. A {\bf 97}, 063809 (2018).

\bibitem{BRJ}B. R. Johnson, M. D. Reed, A. A. Houck, D. I. Schuster, L. S. Bishop, E. Ginossar, J. M. Gambetta, L. DiCarlo, L. Frunzio, S. M. Girvin, and  R. J. Schoelkopf, \textit{Quantum non-demolition detection of single microwave photons in a circuit}, Nature Phys. {\bf 6}, 663 (2010).

\bibitem{waveguide1}M.-A. Lemonde, S. Meesala, A. Sipahigil, M. J. A. Schuetz, M. D. Lukin,  M. Loncar, and P. Rabl, \textit{Phonon networks with silicon-vacancy centers in diamond waveguides}, Phys. Rev. Lett. {\bf 120}, 213603 (2018).

\bibitem{waveguide2}X. Gu, A. F. Kockum, A. Miranowicz, Y.-x. Liu, and F. Nori, \textit{Microwave photonics with superconducting quantum circuits},
Phys. Rep. {\bf 718}, 1 (2017).

\bibitem{EV}{E. V. Stolyarov, \textit{Few-photon Fock-state wave packet interacting with a cavity-atom system in a waveguide: Exact
quantum state dynamics}, Phys. Rev. A {\bf 99}, 023857 (2019)}.



\end{thebibliography}
\end{document}